\documentclass[reprint,twocolumn,aps,prb,showpacs]{revtex4-1}

\usepackage{amsmath}
\usepackage{amssymb}
\usepackage{graphicx}
\usepackage{dcolumn}
\usepackage{bm}
\usepackage[colorlinks=true, allcolors=blue]{hyperref}
\usepackage{epstopdf}
\usepackage[utf8]{inputenc}
\usepackage{amssymb}

\begin{document}

\title{Low-temperature T$^{2}$ resistivity in the underdoped pseudogap phase versus
T-linear resistivity in the overdoped strange-metal phase of cuprate superconductors}

\author{Xingyu Ma$^{1}$, Minghuan Zeng$^{1}$, Huaiming Guo$^{2}$, and Shiping
Feng$^{1}$}
\thanks{Corresponding author. E-mail: spfeng@bnu.edu.cn}
%\email{spfeng@bnu.edu.cn}

\affiliation{$^{1}$Department of Physics, Beijing Normal University, Beijing 100875,
China}

\affiliation{$^{2}$School of Physics, Beihang University, Beijing 100191, China}

%\date{}

\begin{abstract}
The transport experiments demonstrate a dramatic switch from the low-temperature
linear in temperature (T-linear) resistivity in the overdoped strange-metal phase
of cuprate superconductors to the low-temperature quadratic in temperature
(T-quadratic) resistivity in the underdoped pseudogap phase, however, a consensus
on the origin of this unusual switch is still lacking. Here the resistivity in the
underdoped pseudogap phase of cuprate superconductors is investigated using the
Boltzmann transport equation. The resistivity originates from the electron umklapp
scattering mediated by the spin excitation, however, the dominant contribution
mainly comes from {\it the antinodal umklapp scattering}. In particular, a
{\it low temperature} $T_{\rm scale}$ scales with $\Delta^{2}_{p}$ in the
underdoped regime due to the opening of a momentum dependent spin pseudogap, where
$\Delta_{p}$ is the minimal umklapp vector at the antinode. Moreover, this
$T_{\rm scale}$ decreases with the increase of doping in the underdoped regime,
and then is reduced to a {\it very low temperature} in the overdoped regime. In
the underdoped regime, the resistivity is T-quadratic at the low temperatures
below $T_{\rm scale}$, where the strength of the T-quadratic resistivity weakens
as the doping is raised. However, in the overdoped regime, the resistivity is
T-linear at the low temperatures above $T_{\rm scale}$. The results in this paper
together with the recent study on the resistivity in the overdoped regime
therefore show that the electron umklapp scattering from a spin excitation
responsible for the low-temperature T-linear resistivity in the overdoped regime
naturally produces the low-temperature T-quadratic resistivity in the underdoped
regime resulting from the opening of a momentum dependent spin pseudogap.
\end{abstract}


\maketitle

\section{Introduction}\label{Introduction}

It has become very clear that cuprate superconductors are among the most complicated
systems studied in condensed matter physics
\cite{Bednorz86,Damascelli03,Campuzano04,Fink07,Deutscher05,Devereaux07,Fischer07}.
The complications stem mainly from a fact that the parent compound of cuprate
superconductors is an antiferromagnetic (AF) insulator
\cite{Bednorz86,Damascelli03,Campuzano04,Fink07,Deutscher05,Devereaux07,Fischer07}.
Superconductivity then is achieved by chemically introducing charge carriers to
this AF insulator, which leads to that the physical properties mainly depend on the
extent of doping, and the regimes have been classified into the underdoped,
optimally doped, and overdoped, respectively. After intensive studies about four
decades, a substantial amount of reliable and reproducible data has been
accumulated by using many probes
\cite{Damascelli03,Campuzano04,Fink07,Deutscher05,Devereaux07,Fischer07,Hussey08,Kastner98,Timusk99,Hufner08,Vishik18,Fujita12},
which reveals that the most of the unusual features of cuprate superconductors are
observed in the normal-state. In particular, below a pseudogap crossover temperature
$T^{*}$, which can be well above the superconducting (SC) transition temperature
$T_{\rm c}$ in the underdoped regime, the physical response can be well interpreted
in terms of the formation of a pseudogap by which it means a large suppression of
the electronic density of states on the electron Fermi surface (EFS)
\cite{Damascelli03,Campuzano04,Fink07,Deutscher05,Devereaux07,Fischer07,Hussey08,Kastner98,Timusk99,Hufner08,Vishik18,Fujita12}.
This is why in the underdoped regime, the phase above $T_{\rm c}$ but below $T^{*}$
is so-called as {\it the pseudogap phase}. The pseudogap in the underdoped regime
was first discovered through Knight-shift of the magnetic susceptibility experiments
\cite{Warren89,Alloul89}, where the pseudogap is some kind of the {\it spin
pseudogap}, meaning a strong reduction in the magnetic susceptibility through the
particle-hole correlation. Subsequently, the presence of the pseudogap was confirmed
by a series of experimental measurements taken with a wide variety of techniques
\cite{Walstedt90,Takigawa91,Loeser96,Norman98,Renner98,Kohsaka08,Puchkov96}.
Moreover, these experiments also show clearly that the pseudogap exists in both the
{\it spin and charge channels}
\cite{Timusk99,Hufner08,Vishik18,Fujita12,Warren89,Alloul89,Walstedt90,Takigawa91,Loeser96,Norman98,Renner98,Kohsaka08,Puchkov96}.
For example, angle-resolved photoemission spectroscopy (ARPES) and scanning tunneling
spectroscopy measure the charge channel \cite{Loeser96,Norman98}, while nuclear
magnetic resonance (NMR) and nuclear quadrupole resonance (NQR) detect the spin
channel\cite{Fujita12,Warren89,Alloul89,Walstedt90,Takigawa91}. On the other hand, in
the overdoped regime, the normal-state is characterized by a number of the anomalous
low-temperature properties
\cite{Damascelli03,Campuzano04,Fink07,Deutscher05,Devereaux07,Fischer07,Hussey08,Kastner98,Timusk99,Hufner08,Vishik18,Fujita12}
in the sense that they do not fit in with the conventional Fermi-liquid theory (FL)
\cite{Schrieffer64,Abrikosov88,Mahan81}. This has led to the normal-state in the
overdoped regime being refereed to as {\it the strange-metal phase}\cite{Keimer15}.

Among the notable characteristics in the normal-state, the most iconic feature has
to be the electrical transport\cite{Hussey08,Kastner98,Timusk99,Hufner08}. In the
overdoped regime, a series of transport measurements revealed a T-linear
resistivity\cite{Keimer15,Hussey23,Legros19,Ayres21,Grisso21}, where (i) the linear
temperature term often extends to low temperatures of a few kelvin; and (ii) the
linear temperature term persists to high temperatures with the same slope all the
way to the lowest temperature. In this paper, we shall not address the case at high
temperature, but instead focus on the low-temperature T-linear resistivity, which
is in a striking contrast to the conventional FL behaviour in the conventional
superconductors\cite{Schrieffer64,Abrikosov88,Mahan81}. However, in the underdoped
regime, the early transport measurements indicated that the opening of the
pseudogap in the spin excitation spectrum leads to an obvious deviation from the
low-temperature T-linear behaviour of the resistivity\cite{Gurvitch87}. In
particular, the subsequent transport experiments observed a reduced resistivity
below $T^{*}$, near the temperature where the spin pseudogap, as seen by NMR
Knight-shift and spin-relaxation rate, opened\cite{Bucher93,Ito93,Nakano94,Ando01}.
Thus the low-temperature resistivity anomaly coincides roughly with $T^{*}$ and
these experimental results suggest that the decrease in the low-temperature
resistivity below
$T^{*}$ is caused by the reduced electron scattering by the spin excitations
resulting from the opening the spin pseudogap\cite{Bucher93,Ito93,Nakano94,Ando01}.
Lately, the low-temperature resistivity was confirmed experimentally in a quadratic
temperature dependence\cite{Ando04a,Ando04,Lee05,Proust08}, as would be expected
from the conventional FL theory\cite{Schrieffer64,Abrikosov88,Mahan81}, for a wide
doping range in the underdoped regime. More specifically, the recent transport
experiments demonstrate a clear and dramatic switch from the low-temperature
T-linear resistivity in the overdoped regime to the purely T-quadratic resistivity
in the underdoped regime\cite{Cooper09,Hussey11,Mirzaei13,Barisic13,Pelc20}, where
the strength of the T-quadratic resistivity decreases with the increase of doping.
In spite of the considerable variation in crystal structures and impurity-scattering
effects among different families of cuprate superconductors, these experimental
results
\cite{Gurvitch87,Bucher93,Ito93,Nakano94,Ando01,Ando04a,Ando04,Lee05,Proust08,Cooper09,Hussey11,Mirzaei13,Barisic13,Pelc20}
also show that as the low-temperature T-linear resistivity in the overdoped regime
\cite{Legros19,Ayres21,Grisso21}, the T-quadratic resistivity also is a universal
feature in the underdoped pseudogap phase. In this case, some crucial questions are
raised: (i) why the low-temperature resistivity exhibits the T-quadratic behaviour
in the underdoped regime, with an exotic crossover to the T-linear behaviour in the
overdoped regime? (ii) whether the pseudogap is correlated with this exotic
crossover or not? (iii) is there a common electron scattering mediated by the same
bosonic mode that is responsible for both the low-temperature T-quadratic
resistivity in the underdoped regime and low-temperature T-linear resistivity in
the overdoped regime?

The theoretical explanation for this exotic crossover along with the underlying
scattering mechanism is clearly a big challenge. For a understanding of the nature
of the low-temperature T-linear resistivity in the overdoped regime, several
scenarios have been proposed
\cite{Varma89,Varma16,Varma20,Damle97,Sachdev11,Zaanen04,Luca07,Zaanen19,Hartnoll22,Lee21}.
However, these scenarios are just as diverse as the mechanism of superconductivity.
In the marginal FL phenomenology\cite{Varma89,Varma16,Varma20}, a single
T-linear scattering rate is posited to account for the T-linear resistivity. In
particular, the T-linear resistivity was interpreted as a consequence of the scale
invariant physics near to the quantum critical point (QCP)
\cite{Varma20,Damle97,Sachdev11}. With the close relation to the physics of QCP, the
T-linear resistivity has been interpreted in terms of the Planckian
dissipation\cite{Zaanen04,Luca07,Zaanen19,Hartnoll22}. Moreover, it has been shown
recently that the low-temperature T-linear resistivity originates from the umklapp
scattering between electrons by the exchange of a critical boson propagator
\cite{Lee21}. These studies\cite{Lee21}
together with many others\cite{Honerkamp01,Hartnoll12} thus suggest that the
electron umklapp scattering is the origin of the low-temperature T-linear
resistivity. On the other hand, it has been argued that if the electron scattering
responsible for the low-temperature T-linear resistivity in the overdoped regime
involves the electron scattering on the spin excitations in the underdoped regime
\cite{Puchkov96}, then the spin pseudogap seen in NMR below $T^{*}$ would naturally
account for a deviation from the low-temperature T-linear behaviour of the
resistivity\cite{Bucher93,Ito93,Nakano94,Ando01}. In particular, the mechanism for
the resistivity in the underdoped regime has been proposed\cite{Hussey03,Rice17},
where the electron umklapp scattering above $T^{*}$ leads to the T-linear
resistivity\cite{Rice17}. As the temperatures fall below $T^{*}$, the pseudogap
opens, and then restricts the available umklapp scattering channels, leading to a
crossover between the T-linear resistivity at the temperature above $T^{*}$ and
T-quadratic form of the scattering from the Fermi pockets at the
temperature below $T^{*}$. Moreover, the low-temperature resistivity has been
studied based on the two-particle self-consistent approach\cite{Bergeron11}, where
it has been shown that the resistivity is linear at low temperatures in the
overdoped regime, however, in the underdoped regime, the resistivity may display
a T-quadratic behaviour at temperatures below $T^{*}$. However, up to now, the
T-quadratic resistivity in the underdoped pseudogap phase and its connection with
the pseudogap is still the subject of much study, and the microscopic origin of the
clear switch from the low-temperature T-linear resistivity in the overdoped
regime to the low-temperature T-quadratic resistivity in the underdoped regime is
hotly debated.

In the very recent study\cite{Ma23}, we have studied the nature of the
low-temperature T-linear resistivity in the overdoped regime. In our scenario, the
scattering rate arises from the umklapp scattering between electrons by the exchange
of the effective spin propagator, where the dominant contribution mainly comes from
{\it the antinodal umklapp scattering}. This umklapp scattering rate scales linearly
with temperature in the low temperatures, which then naturally generates a
low-temperature T-linear resistivity. In this paper, we study the low-temperature
resistivity in the underdoped pseudogap phase along with this line, where in a
striking difference to the case in the overdoped regime, the spin excitation energy
dispersion is anisotropically renormalized due to the opening of a momentum
dependent spin pseudogap. In particular, the density of the spin excitations at
around the antinodal is heavily reduced by the antinodal spin pseudogap. In this
case, a {\it low temperature} $T_{\rm scale}$ scales with $\Delta^{2}_{p}$, where
$\Delta_{p}$ is the minimal umklapp vector at the antinode. Notably, this
{\it low-temperature scale} $T_{\rm scale}$ as a function of doping presents a
similar behavior of the antinodal spin pseudogap crossover temperature, i.e.,
$T_{\rm scale}$ is dropped down with the enhancement of doping in the underdoped
regime, and then is reduced to a {\it very low temperature} in the overdoped regime.
In the underdoped regime, the resistivity is T-quadratic at the low temperatures
below $T_{\rm scale}$, where the T-quadratic resistivity strength weakens as the
doping is raised. However, in the overdoped regime, the resistivity is instead
T-linear in the low temperatures above $T_{\rm scale}$. The present results combined
with the recent results\cite{Ma23} on the low-temperature T-linear resistivity in
the overdoped regime therefore reveal that (i) the electron umklapp scattering from
a spin excitation associated with the antinodes leads to the T-linear resistivity
in the weak coupling overdoped regime; (ii) as this electron umklapp scattering
flows to the strong coupling underdoped regime, the opening of the momentum
dependent spin pseudogap lowers the density of the spin excitations at around the
antinodal region, which reduces the strength of the intense umklapp scattering from
the electronic states into the antinodal region, and therefore leads to the
low-temperature T-quadratic form of the umklapp scattering rate.

This paper is organized as follows. In the next section, we first show that (i) the
antinodal spin pseudogap decreases with the increase of doping in the underdoped
regime, and then it abruptly vanishes at around the optimal doping; (ii) the
electronic density of state at around the antinodal region is gapped out by the
normal-state pseudogap, and then the closed EFS contour is truncated to a set of
four disconnected Fermi arcs centered at around the nodal region. Following this
reconstructed EFS, the scattering rate originated from the umklapp scattering
between electrons by the exchange of the full effective spin propagator is derived
using the Boltzmann transport equation. The discussions of the quantitative
characteristics of the low-temperature resistivity in the underdoped pseudogap phase
are given in Section \ref{electron-resistivity}, where we show that the scattering
rate in the whole doping regime is predominantly governed by the antinodal umklapp
scattering. In Section \ref{summary}, we close the presentation with a summary and
discussion of the main results. In Appendix \ref{charge-spin-propagator}, we present
the details for the derivation of the full spin and full electron propagators.

\section{Methodology}\label{Formalism}

\subsection{Low-energy effective $t$-$J$ model}\label{model-constraint}

The common element in the layered crystal structure of cuprate superconductors is
the square-lattice copper-oxide layer\cite{Bednorz86}, and then the unusual features
mainly come from the strongly correlated motion of the electrons in the copper-oxide
layer\cite{Cooper94,Nakamura93,Hou94,Takenaka94}. Quickly after the discovery of
superconductivity in cuprate superconductors, Anderson\cite{Anderson87} recognized
that the essential physics of the doped copper-oxide layer can be modeled with the
low-energy effective $t$-$J$ model,
\begin{eqnarray}\label{tJ-model}
H&=&-t\sum_{\langle l\hat{\eta}\rangle\sigma}C^{\dagger}_{l\sigma}
C_{l+\hat{\eta}\sigma}+t'\sum_{\langle l\hat{\tau}\rangle\sigma}
C^{\dagger}_{l\sigma}C_{l+\hat{\tau}\sigma}+\mu\sum_{l\sigma}C^{\dagger}_{l\sigma}
C_{l\sigma}\nonumber\\
&+&J\sum_{\langle l\hat{\eta}\rangle}{\bf S}_{l}\cdot {\bf S}_{l+\hat{\eta}},~~~~
\end{eqnarray}
where $C^{\dagger}_{l\sigma}$ ($C_{l\sigma}$) is electron operator that creates
(annihilates) an electron on site $l$ with spin $\sigma$, ${\bf S}_{l}$ is the spin
operator with its components $S_{l}^{x}$,
$S_{l}^{y}$, and $S_{l}^{z}$, and $\mu$ is the chemical potential. This $t$-$J$
model (\ref{tJ-model}) describes a competition between the kinetic energy and
magnetic energy, and the kinetic energy includes the nearest-neighbor (NN) hopping
with the hopping integral $t$ and next NN hopping with the hopping integral
$t'$, while the magnetic energy is described by an AF
Heisenberg term with the NN coupling $J$. $\langle l\hat{\eta}\rangle$
($\langle l\hat{\tau}\rangle$) indicates that $l$ runs over all sites, and for each
$l$, over its NN sites $\hat{\eta}$ (next NN sites $\hat{\tau}$). The parameters in
this paper are chosen as $t/J=2.5$ and $t'/t=0.3$ as the recent discussions
\cite{Ma23}. Moreover, we use a notation in which the magnetic coupling $J$ and the
lattice constant of the square lattice are set to the energy and length units,
respectively. However, when necessary to compare with the experimental data, we
take $J=100$meV, which is the typical value of cuprate superconductors.

The $t$-$J$ model (\ref{tJ-model}) is supplemented by a crucial on-site local
constraint $\sum_{\sigma}C^{\dagger}_{l\sigma}C_{l\sigma}\leq 1$ that the double
occupancy of a site by two electrons of opposite spins is not allowed
\cite{Anderson87}, while the strong electron correlation in the system manifests
itself by this no double electron occupancy local constraint
\cite{Yu92,Lee06,Edegger07,Spalek22,Zhang93}. However, the no double electron
occupancy also leads to the difficulty for the studying of the $t$-$J$ model
(\ref{tJ-model}). For a proper treatment of this no double electron occupancy local
constraint, the fermion-spin transformation\cite{Feng0494,Feng15} has been
established, where the physics of no double electron occupancy is taken into account
by representing the constrained electron as,
\begin{eqnarray}\label{CSSFS}
C_{l\uparrow}=h^{\dagger}_{l\uparrow}S^{-}_{l},~~~~
C_{l\downarrow}=h^{\dagger}_{l\downarrow}S^{+}_{l},
\end{eqnarray}
where the $U(1)$ gauge invariant spin-raising (spin-lowering) operator $S^{+}_{l}$
($S^{-}_{l}$) carries spin index of the constrained electron, and therefore the
collective mode from this spin degree of freedom of the constrained electron is
interpreted as the spin excitation responsible for the spin dynamics of the system,
and the $U(1)$ gauge invariant spinful fermion operator
$h^{\dagger}_{l\sigma}=e^{i\Phi_{l\sigma}}h^{\dagger}_{l}$
($h_{l\sigma}=e^{-i\Phi_{l\sigma}}h_{l}$) creates (annihilates) a charge carrier on
site $l$, and therefore represents the charge degree of freedom of the constrained
electron together with some effects of spin configuration rearrangements due to the
presence of the doped charge carrier itself, while the charge carrier and localized
spin recombine to form the physical electron responsible for the electronic
properties.

\subsection{Pseudogap in charge and spin channels}\label{spin-charge-pseudogaps}

In the past three decades, a series of experiments from NMR, NQR, and the inelastic
neutron scattering (INS) measurements
\cite{Fujita12,Birgeneau89,Fong95,Yamada98,Arai99,Bourges00,He01,Tranquada04,Bourges05}
has provided an intrinsic connection between the electron pairing mechanism and spin
excitations in cuprate superconductors, where a key question is whether the spin
excitation can mediate electron pairing in analogy to the phonon-mediate pairing
mechanism in the conventional superconductors\cite{Schrieffer64}. Starting from the
$t$-$J$ model in the fermion-spin representation, the kinetic-energy-driven
superconductivity\cite{Feng15,Feng0306,Feng12,Feng15a} has been developed, where the
d-wave charge-carrier pairing state is generated by the charge-carrier interaction
directly from the kinetic energy of the $t$-$J$ model by the exchange of the spin
excitation. However, the d-wave electron
pairs originated from this d-wave charge-carrier pairing state are due to the
charge-spin recombination\cite{Feng15a}, and these electron pairs condensation
reveals the d-wave SC-state. This kinetic-energy-driven SC mechanism is purely
electronic without phonon, since the glue to hold the constrained electron pairs
together is {\it the spin excitation, the collective mode from the spin degree of
freedom of the constrained electron itself}. Moreover, this electron paring state
mediated by the spin excitation in a way is in turn strongly influenced by the
electron coherence, which leads to that the maximal $T_{\rm c}$ occurs at the
optimal doping, and then decreases in both the underdoped and overdoped regimes.
Following these previous discussions\cite{Feng15,Feng0306,Feng12,Feng15a}, the
full charge-carrier propagator in the normal-state can be derived as [see
Appendix \ref{charge-spin-propagator}],
\begin{eqnarray}\label{FCCGF}
g({\bf k},\omega)={1\over\omega-\xi_{\bf k}
-\Sigma^{({\rm h})}_{\rm ph}({\bf k},\omega)},
\end{eqnarray}
where
$\xi_{\bf k}=4t\chi_{1}\gamma_{{\bf k}}-4t'\chi_{2}\gamma_{{\bf k}}'-\mu_{\rm h}$
is the mean-filed (MF) charge-carrier energy dispersion, with
$\gamma_{\bf k}=({\rm cos}k_{x}+{\rm cos} k_{y})/2$,
$\gamma_{\bf k}'={\rm cos}k_{x}{\rm cos}k_{y}$, and the spin correlation functions
$\chi_{1}=\langle S^{+}_{l}S^{-}_{l+\hat{\eta}}\rangle$ and
$\chi_{2}=\langle S_{l}^{+}S_{l+\hat{\tau}}^{-}\rangle$, while the charge-carrier
normal self-energy $\Sigma^{({\rm h})}_{\rm ph}({\bf k},\omega)$ is obtained
explicitly in Eq. (\ref{HESE-1}) in Appendix \ref{charge-spin-propagator}.

In the framework of the kinetic-energy-driven superconductivity
\cite{Feng15,Feng0306,Feng12,Feng15a}, the charge-carrier pseudogap forms due to
the strong coupling between charge and spin degrees of freedom of the constrained
electron. To explore the nature of this charge-carrier pseudogap more clearly, we
rewrite the charge-carrier normal self-energy in Eq. (\ref{FCCGF}) as,
\begin{eqnarray}\label{self-energy-2}
\Sigma^{({\rm h})}_{\rm ph}(\bf k,\omega)&\approx&
{[\bar{\Delta}^{({\rm h})}_{\rm pg}({\bf k})]^{2}\over\omega-\xi_{0{\bf k}}},
\end{eqnarray}
where $\xi_{0{\bf k}}=L^{({\rm h})}_{2}({\bf k})/L^{({\rm h})}_{1}({\bf k})$ is
the energy spectrum of $\Sigma^{({\rm h})}_{\rm ph}(\bf k,\omega)$,
and $\bar{\Delta}^{({\rm h})}_{\rm pg}({\bf k})=L^{({\rm h})}_{2}({\bf k})
/\sqrt{L^{({\rm h})}_{1}({\bf k})}$ is refereed to as the momentum dependence of the
charge-carrier pseudogap, since it plays a role of the suppression of the low-energy
spectral weight of the charge-carrier excitation spectrum, with the functions
$L^{({\rm h})}_{1}({\bf k})=-\Sigma^{({\rm h})}_{\rm pho}({\bf k},\omega=0)$ and
$L^{({\rm h})}_{2}({\bf k})=-\Sigma^{({\rm h})}_{\rm ph}({\bf k},\omega=0)$, while
$\Sigma^{({\rm h})}_{\rm ph}({\bf k},\omega=0)$ and the antisymmetric part
$\Sigma^{({\rm h})}_{\rm pho}({\bf k},\omega)$ of the charge-carrier normal
self-energy are obtained directly from $\Sigma^{({\rm h})}_{\rm ph}({\bf k},\omega)$
in Eq. (\ref{HESE-1}).
\begin{figure}[h!]
\includegraphics[scale=0.70]{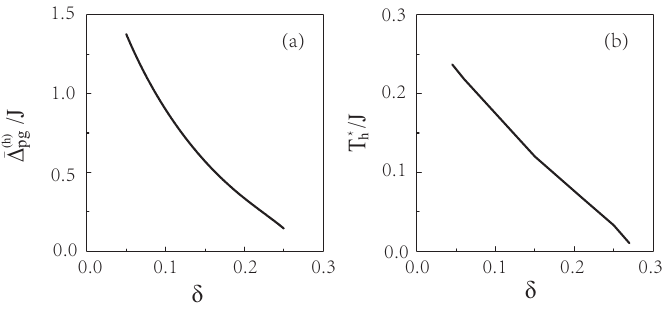}
\caption{(a) The charge-carrier pseudogap at temperature $T=0.002J$ and (b) the
charge-carrier pseudogap crossover temperature as a function of doping taken from
Ref. \onlinecite{Feng12}.\label{CCPG-T-doping}}
\end{figure}
Moreover, the sharp peaks appearing at low-temperature in
$\Sigma^{({\rm h})}_{\rm ph}(\bf k,\omega)$ and the related quantities are actually
a $\delta$-function, which are broadened by a small damping employed in the
numerical calculation for a finite lattice\cite{Brinckmann01,Restrepo23}. As the
same approach of the numerical calculation carried out in Ref. \onlinecite{Ma23},
the calculation in this paper for $\Sigma^{({\rm h})}_{\rm ph}(\bf k,\omega)$ and
the related quantities is performed numerically on a $160\times 160$ lattice in
momentum space, where the infinitesimal $i0_{+}\rightarrow i\Gamma$ is replaced by
a small damping $\Gamma=0.05J$.

For a convenience in the following discussions, the charge-carrier pseudogap
\cite{Feng15,Feng12} $\bar{\Delta}^{({\rm h})}_{\rm pg}$ at temperature $T=0.002J$
as a function of doping is replotted in Fig. \ref{CCPG-T-doping}a, where the
relatively large $\bar{\Delta}^{({\rm h})}_{\rm pg}$ appears in the underdoped
regime, and then it weakens as the optimal doping is approached. However, a quite
weak $\bar{\Delta}^{({\rm h})}_{\rm pg}$ is still present at around the optimal
doping, but it disappears at the heavily overdoped region. Moreover, at a given
doping concentration, this $\bar{\Delta}^{({\rm h})}_{\rm pg}$ is identified with
a crossover with a charge-carrier pseudogap crossover temperature $T^{*}_{\rm h}$
rather than a phase transition. To see the evolution of $T^{*}_{\rm h}$ with doping
more clearly, $T^{*}_{\rm h}$ as a function of doping is also replotted in
Fig. \ref{CCPG-T-doping}b, where in conformity with the doping dependence of
$\bar{\Delta}^{({\rm h})}_{\rm pg}$ in Fig. \ref{CCPG-T-doping}a, $T^{*}_{\rm h}$ is
relatively high at the slight underdoping, and then it smoothly decreases with the
increase of doping in the underdoped regime, eventually terminating at the heavily
overdoped region\cite{Feng12,Feng15}. More importantly, as the electron pairing
state originated from the charge-carrier pairing state are due to the charge-spin
recombination, the normal-state pseudogap state originated from the charge-carrier
pseudogap state is also due to the charge-spin recombination\cite{Feng15a}, and then
the anomalous properties associated with the formation of the normal-state pseudogap
are explained in a natural way
\cite{Damascelli03,Campuzano04,Fink07,Deutscher05,Devereaux07,Fischer07,Hussey08,Kastner98,Timusk99,Hufner08,Vishik18,Fujita12}.
We will return to the discussion of the normal-state pseudogap towards next
subsection \ref{Normal-state-pseudogap}.

On the other hand, in the fermion-spin theory\cite{Feng0494,Feng15}, the scattering
of spins due to the charge-carrier fluctuation dominates the spin dynamics. In this
case, the dynamical spin response of cuprate superconductors has been discussed from
the SC-state to the normal-state\cite{Kuang15,Yuan01,Feng98}, where the full spin
propagator in the normal-state has been derived as
[see Appendix \ref{charge-spin-propagator}],
\begin{eqnarray}\label{FSGF}
D({\bf k},\omega)&=&{B_{\bf k}\over \omega^{2}-\omega^{2}_{\bf k}
-B_{\bf k}\Sigma^{({\rm s})}_{\rm ph}({\bf k},\omega)},~~~~~
\end{eqnarray}
with the MF spin excitation energy dispersion $\omega_{\bf k}$ and the corresponding
weight function $B_{\bf k}$ of the MF spin excitation spectrum that have been
obtained explicitly in the previous works\cite{Feng15,Kuang15}, while the spin
self-energy in the normal-state obtained in terms of the collective charge-carrier
mode in the particle-hole channel that is given in Eq. (\ref{SSFN}) in
Appendix \ref{charge-spin-propagator}.

For a better understanding of the nonconventional features of the spin pseudogap,
the above spin self-energy in Eq. (\ref{FSGF}) can be also rewritten as,
\begin{equation}\label{SSFN-spin-gap}
\Sigma^{({\rm s})}_{\rm ph}({\bf k},\omega)
\approx {B_{\bf k}[\bar{\Delta}^{({\rm s})}_{\rm pg}({\bf k})]^{2}\over \omega^{2}
-\omega^{2}_{0{\bf k}}},
\end{equation}
where $\omega^{2}_{0{\bf k}}=L^{({\rm s})}_{1{\bf k}}/L^{({\rm s})}_{2{\bf k}}$ is
the energy spectrum of $\Sigma^{({\rm s})}_{\rm ph}({\bf k},\omega)$, while
$[\bar{\Delta}^{({\rm s})}_{\rm pg}({\bf k})]^{2}=[L^{({\rm s})}_{1{\bf k}}]^{2}
/B_{\bf k}L^{({\rm s})}_{2{\bf k}}$ is identified as being a region of the spin
self-energy in which $\bar{\Delta}^{({\rm s})}_{\rm pg}({\bf k})$ anisotropically
reduces the density of the spin excitation, and in this sense,
$\bar{\Delta}^{({\rm s})}_{\rm pg}({\bf k})$ is refereed to as the spin pseudogap.
The functions $L^{({\rm s})}_{1{\bf k}}$ and $L^{({\rm s})}_{2{\bf k}}$ is derived
directly from the spin self-energy $\Sigma^{({\rm s})}_{\rm ph}({\bf k},\omega)$ in
Eq. (\ref{SSFN}) as,
\begin{subequations}
\begin{eqnarray}
L^{({\rm s})}_{1{\bf k}}&=&-{B_{\bf k}\over N^{2}}\sum_{\bf pq}
\Omega^{({\rm s})}_{{\bf k}{\bf p}{\bf q}}{F^{\rm (s)}({\bf k},{\bf p},{\bf q})\over
[\omega_{{\bf q}+{\bf k}}-(\bar{\xi}_{{\bf p}+{\bf q}}-\bar{\xi}_{{\bf p}})]^{2}},
~~~~~\\
L^{({\rm s})}_{2{\bf k}}&=&-{B_{\bf k}\over N^{2}}\sum_{\bf pq}
\Omega^{({\rm s})}_{{\bf k}{\bf p}{\bf q}}{F^{\rm (s)}({\bf k},{\bf p},{\bf q})\over
[\omega_{{\bf q}+{\bf k}}-(\bar{\xi}_{{\bf p}+{\bf q}}-\bar{\xi}_{{\bf p}})]^{4}},
~~~~~~~~
\end{eqnarray}
\end{subequations}
respectively, where the renormalized charge-carrier energy dispersion
$\bar{\xi}_{\bf k}=Z^{({\rm h})}_{\rm F}\xi_{\bf k}$, with the charge-carrier
coherent weight $Z^{({\rm h})-1}_{\rm F}
=1-{\rm Re}\Sigma^{({\rm h})}_{\rm pho}({\bf k},0)\mid_{{\bf k}=[\pi,0]}$, while
the function $F^{\rm (s)}({\bf k},{\bf p},{\bf q})$ and vertex function
$\Omega^{({\rm s})}_{{\bf k}{\bf p}{\bf q}}$ are presented in
Appendix \ref{charge-spin-propagator}. Substituting the above spin
self-energy (\ref{SSFN-spin-gap}) into Eq. (\ref{FSGF}), the full spin propagator
in the normal-state can be derived as,
\begin{eqnarray}\label{FSGF-spin-gap}
D({\bf k},\omega)={\bar{B}_{1{\bf k}}\over\omega^{2}-\bar{\omega}^{2}_{1{\bf k}}}
+{\bar{B}_{2{\bf k}}\over\omega^{2}-\bar{\omega}^{2}_{2{\bf k}}}
=\sum_{\alpha=1,2}{\bar{B}_{\alpha{\bf k}}\over\omega^{2}
-\bar{\omega}^{2}_{\alpha{\bf k}}},~~~~~~
\end{eqnarray}
with the renormalized spin excitation energy dispersions,
\begin{subequations}\label{FSEED}
\begin{eqnarray}
\bar{\omega}^{2}_{1{\bf k}}&=&{1\over 2}\left [\omega^{2}_{\bf k}
+\omega^{2}_{0{\bf k}}+\sqrt{(\omega^{2}_{\bf k}-\omega^{2}_{0{\bf k}})^{2}
+4B^{2}_{\bf k}[\bar{\Delta}^{({\rm s})}_{\rm pg}({\bf k})]^{2}}\right ],\nonumber\\
~~~~~~ \\
\bar{\omega}^{2}_{2{\bf k}}&=&{1\over 2}\left [\omega^{2}_{\bf k}
+\omega^{2}_{0{\bf k}}-\sqrt{(\omega^{2}_{\bf k}-\omega^{2}_{0{\bf k}})^{2}
+4B^{2}_{\bf k}[\bar{\Delta}^{({\rm s})}_{\rm pg}({\bf k})]^{2}}\right ],\nonumber\\
~~~~~~
\end{eqnarray}
\end{subequations}
and the corresponding weight functions of the full spin excitation spectrum,
\begin{figure}[h!]
\includegraphics[scale=0.67]{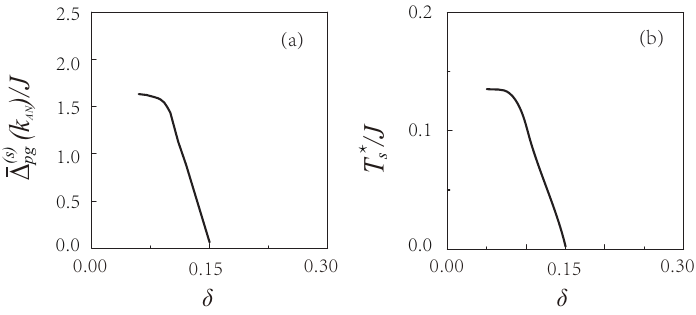}
\caption{(a) The spin pseudogap $\bar{\Delta}^{({\rm s})}_{\rm pg}({\bf k}_{\rm AN})$
at the antinode for temperature $T=0.002J$ and (b) the corresponding antinodal
spin-pseudogap crossover temperature $T^{*}_{\rm s}$ as a function of doping, where
${\bf k}_{\rm AN}$ is the wave vector at the antinode.\label{spin-pseudogap-doping}}
\end{figure}
\begin{subequations}\label{weight-functions}
\begin{eqnarray}
\bar{B}_{1{\bf k}}&=& {1\over 2}B_{\bf k}\left ({\omega^{2}_{\bf k}
-\omega^{2}_{0{\bf k}}\over \sqrt{(\omega^{2}_{\bf k}-\omega^{2}_{0{\bf k}})^{2}
+4B^{2}_{\bf k}[\bar{\Delta}^{({\rm s})}_{\rm pg}({\bf k})}]^{4}}+1\right ),~~~~\\
\bar{B}_{2{\bf k}}&=& -{1\over 2}B_{\bf k}\left ({\omega^{2}_{\bf k}
-\omega^{2}_{0{\bf k}}\over \sqrt{(\omega^{2}_{\bf k}-\omega^{2}_{0{\bf k}})^{2}
+4B^{2}_{\bf k}[\bar{\Delta}^{({\rm s})}_{\rm pg}({\bf k})}]^{4}}-1\right ).~~~~~~~~
\end{eqnarray}
\end{subequations}
It should be emphasized that the equation (\ref{SSFN-spin-gap}) is an identity only
in the case of $\omega =0$, however, as in the case of the charge-carrier pseudogap
\cite{Feng15,Feng12}, it is a proper approximation for the low-energy case of
$\omega\neq 0$.

The momentum dependence of the spin pseudogap
$\bar{\Delta}^{({\rm s})}_{\rm pg}({\bf k})$ in Eq. (\ref{SSFN-spin-gap}) evolves
strongly with doping and temperature, as that in the charge-carrier pseudogap
\cite{Feng15,Feng12}. On the other hand, we\cite{Ma23} have shown that the umklapp
scattering between electrons by the exchange of the MF effective spin propagator
can give a consistent description of the low-temperature T-linear resistivity in
the overdoped strange-metal phase, where the dominant contribution to the scattering
rate mainly comes from the antinodal umklapp scattering. In this case, we focus on
the exotic features of the doping and temperature dependent spin pseudogap
$\bar{\Delta}^{({\rm s})}_{\rm pg}({\bf k})$ at around the antinodal region for a
convenience in the following discussions of the low-temperature resistivity in the
underdoped pseudogap phase.
In Fig. \ref{spin-pseudogap-doping}a,
we plot $\bar{\Delta}^{({\rm s})}_{\rm pg}({\bf k}_{\rm AN})$ at the antinode as a
function of doping with temperature $T=0.002J$, where ${\bf k}_{\rm AN}$ is the wave
vector at the antinode of the Brillouin zone (BZ). Apparently, this antinodal spin
pseudogap is insensitive to the doping in the slightly underdoped region, and then
it falls off rapidly as the doping is grown in the heavily underdoped region.
To our big surprise, this antinodal spin pseudogap abruptly disappears at around the
optimal doping. This antinodal spin pseudogap formation in the underdoped regime
lowers the density of the spin excitations in response to the intense electron
umklapp scattering from the spin excitations associated with the antinodes. However,
in the overdoped regime, the main properties of the antinodal spin excitations can
be well described by the MF spin propagator. Furthermore, for a given
doping, this antinodal spin pseudogap vanishes when temperature reaches the
antinodal spin-pseudogap crossover temperature $T^{*}_{\rm s}$. To see this doping
dependence of $T^{*}_{\rm s}$ more clearly, we plot $T^{*}_{\rm s}$ as a function of
doping in Fig. \ref{spin-pseudogap-doping}b, where in corresponding to the result of
$\bar{\Delta}^{({\rm s})}_{\rm pg}({\bf k}_{\rm AN})$ in
Fig. \ref{spin-pseudogap-doping}a, $T^{*}_{\rm s}$ is relatively high in the slightly
underdoped region, and then it decreases rapidly when the doping is increased in the
heavily underdoped region, eventually disappearing at around the optimal doping. It
should be noted that the disappearance of the antinodal spin pseudogap at around the
optimal doping does not indicate the existence of QCP at around the optimal doping.
This follows from a basic fact that the experimental observations in the optimally
doped and overdoped regimes show that although the strength of the T-linear
resistivity gradually diminishes as a function of doping, the low-temperature
T-linear resistivity retains a finite value up to the edge of the SC dome
\cite{Keimer15,Hussey23,Legros19,Ayres21,Grisso21}, which can be described by the
momentum relaxation due to the umklapp scattering between electrons by the exchange
of the MF effective spin propagator\cite{Ma23},
however, it may be a challenging issue for the usual picture of a single QCP doping
associated with the strange-metal phase\cite{Varma20,Damle97,Sachdev11}.

Based on the full spin propagator (\ref{FSGF-spin-gap}), the dynamical spin response
of cuprate superconductors in the
normal-state has been investigated\cite{Kuang15,Yuan01,Feng98}, where the obtained
results show the existence of damped but well-defined dispersive spin excitations in
the whole doping phase diagram. In particular, the spectral weight of the spin
excitation spectrum at around the antinodal region is strongly suppressed by the
antinodal spin pseudogap\cite{Kuang15}. Moreover, the low-energy spin fluctuation is
dominated by the process from the mobile charge carriers, while the high-energy spin
excitation on the other hand retains roughly constant energy as a function of
doping, with spectral weight and dispersion relation comparable to those in the
corresponding SC-state. All these results are qualitatively consistent with the
experimental observations
\cite{Fujita12,Birgeneau89,Fong95,Yamada98,Arai99,Bourges00,He01,Tranquada04,Bourges05}.

\subsection{Electronic structure in the underdoped pseudogap state}
\label{Normal-state-pseudogap}

For the understanding of the nature of the electronic structure in cuprate
superconductors, it is needed to derive the electron propagator, which is
characterized by the full charge-spin recombination\cite{Feng15a}. Following the
previous discussions\cite{Feng15a}, the full electron propagator in the
normal-state can be derived as [see Appendix \ref{charge-spin-propagator}],
\begin{eqnarray}\label{EGF}
G({\bf k},\omega)={1\over \omega-\varepsilon_{\bf k}
-\Sigma_{\rm ph}({\bf k},\omega)},
\end{eqnarray}
where $\varepsilon_{\bf k}=-4t\gamma_{\bf k}+4t'\gamma_{\bf k}'+\mu$ is the electron
energy dispersion in the tight-binding approximation, while the electron normal
self-energy $\Sigma_{\rm ph}({\bf k},\omega)$ is given in Eq. (\ref{ESE-5}) in
Appendix \ref{charge-spin-propagator}.

The electron normal self-energy in Eq. (\ref{EGF}) originated from the charge-carrier
normal self-energy in Eq. (\ref{FCCGF}) is due to the charge-spin recombination
\cite{Feng15a}, indicating that the normal-state pseudogap state originated from the
charge-carrier pseudogap state is due to the charge-spin recombination. To see the
nature of the normal-state pseudogap more clearly, the above electron normal
self-energy in Eq. (\ref{EGF}) can be reexpressed as,
\begin{eqnarray}\label{EPG}
\Sigma_{\rm ph}({\bf k},\omega)&\approx& {[\bar{\Delta}_{\rm PG}({\bf k})]^{2}\over
\omega -\varepsilon_{0{\bf k}}},
\end{eqnarray}
where $\varepsilon_{0{\bf k}}=L_{2{\bf k}}/L_{1{\bf k}}$ is the energy spectrum of
$\Sigma_{\rm ph}({\bf k},\omega)$, and
$\bar{\Delta}^{2}_{\rm PG}({\bf k})=L^{2}_{2{\bf k}}/L_{1{\bf k}}$ is the
normal-state pseudogap, with the functions
$L_{1}({\bf k})=-\Sigma_{\rm pho}({\bf k},\omega=0)$ and
$L_{2} ({\bf k})=-\Sigma_{\rm ph}({\bf k},\omega=0)$, while
the antisymmetric part
$\Sigma_{\rm pho}({\bf k},\omega)$ of the electron normal self-energy can be
obtained directly from the electron normal self-energy
$\Sigma_{\rm ph}({\bf k},\omega)$ in Eq. (\ref{ESE-5}). This
normal-state pseudogap $\bar{\Delta}_{\rm PG}({\bf k})$ is also identified as being
a region of the electron normal self-energy in which the normal-state pseudogap
anisotropically suppresses the electronic density of states on EFS. Since the
normal-state pseudogap state is induced by the charge-carrier pseudogap state, the
normal-state pseudogap $\bar{\Delta}_{\rm PG}({\bf k})$ [the normal-state pseudogap
crossover temperature $T^{*}$] as a function of doping presents a similar behavior
of the charge-carrier pseudogap $\bar{\Delta}^{({\rm h})}_{\rm pg}({\bf k})$ [the
charge-carrier pseudogap crossover temperature $T^{*}_{\rm h}$]\cite{Feng15,Feng12}.
This normal-state pseudogap crossover temperature $T^{*}$ is actually a crossover
line below which a novel electronic state emerges, as exemplified by the presence of
the Fermi arcs, the competing electronic orders, etc., and then the unconventional
features of this novel electronic state can be well interpreted in terms of the
formation of the normal-state pseudogap
\cite{Damascelli03,Campuzano04,Fink07,Deutscher05,Devereaux07,Fischer07,Hussey08,Kastner98,Timusk99,Hufner08,Vishik18}.

The measured energy and momentum distribution curves can be depicted by the electron
excitation spectrum
\cite{Dessau91,Randeria95,Fedorov99,Campuzano99,DMou17,Bogdanov00,Kaminski01,Johnson01,Iwasawa08,Plumb13},
\begin{equation}\label{EXS-ED}
I({\bf k},\omega)\propto n_{\rm F}(\omega) A({\bf k},\omega),
\end{equation}
with the electron spectral function,
\begin{equation}\label{ESF}
A({\bf k},\omega)={1\over\pi}{{\rm Im}\Sigma_{\rm ph}({\bf k},\omega)\over
[\omega-\varepsilon_{\bf k}-{\rm Re}\Sigma_{\rm ph}({\bf k},\omega)]^{2}
+[{\rm Im}\Sigma_{\rm ph}({\bf k},\omega)]^{2}},~~~
\end{equation}
where ${\rm Re}\Sigma_{\rm ph}({\bf k},\omega)$ and
${\rm Im}\Sigma_{\rm ph}({\bf k},\omega)$ are the real and imaginary parts of
$\Sigma_{\rm ph}({\bf k},\omega)$, respectively. The
electrons are renormalzed due to the electron scattering mediated by the spin
excitation and then they acquire a finite lifetime.

The EFS topology is known to be crucial in the understanding of the unconventional
superconductivity in cuprate superconductors
\cite{Damascelli03,Campuzano04,Fink07,Deutscher05,Devereaux07,Fischer07} and the
related anomalous normal-state properties
\cite{Hussey08,Kastner98,Timusk99,Hufner08,Vishik18}. The ARPES measured EFS can be
described theoretically by the intensity map of the electron excitation spectrum
(\ref{EXS-ED}) at zero energy $\omega=0$, where the locations of the EFS continuous
contour in momentum space is determined directly by the poles of the electron
spectral function ({\ref{ESF}}),
$\varepsilon_{\bf k}+{\rm Re}\Sigma_{\rm ph}({\bf k},0)=\bar{\varepsilon}_{\bf k}=0$,
with the renormalized electron energy dispersion
$\bar{\varepsilon}_{\bf k}=Z_{\rm F}\varepsilon_{\bf k}$ and the single-particle
coherent weight
$Z^{-1}_{\rm F}=1-{\rm Re}\Sigma_{\rm pho}({\bf k},0)\mid_{{\bf k}=[\pi,0]}$.
However, the strong redistribution of the spectral weight on EFS is mainly governed
by the momentum dependence of the normal-state pseudogap
$\bar{\Delta}_{\rm PG}({\bf k})$ [then the single-electron scattering rate
$\Gamma_{\bf k}(\omega)={\rm Im}\Sigma_{\rm ph}({\bf k},\omega)
=\pi [\bar{\Delta}_{\rm PG}({\bf k})]^{2}\delta(\omega+\varepsilon_{0{\bf k}})]$.
\begin{figure}[h!]
\centering
\includegraphics[scale=0.85]{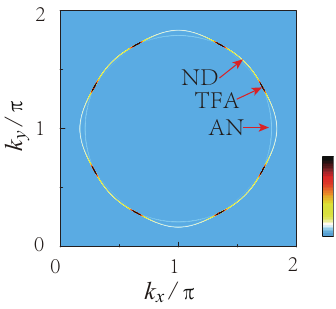}
\caption{(Color online) Electron Fermi surface at $\delta=0.09$ with $T=0.002J$,
where the Brillouin zone center has been shifted by [$\pi,\pi$], and AN, TFA, and ND
denote the antinode, tip of the Fermi arc, and node, respectively. \label{EFS-map}}
\end{figure}
For a convenience in the following discussions of the resistivity in the
underdoped pseudogap phase, the EFS map\cite{Feng16,Liu21,Cao21,Zeng22} at doping
$\delta=0.09$ with temperature $T=0.002J$ is replotted in Fig. \ref{EFS-map}, where
the BZ center has been shifted by [$\pi,\pi$], and AN, TFA, and ND
indicate the antinode, tip of the Fermi arc, and node, respectively. It thus shows
clearly that the antinodal region of EFS becomes partially gapped, leading to that
EFS consists, not of a closed contour, but only of four disconnected Fermi arcs
centered around the nodes, in qualitative agreement with the experimental results
\cite{Loeser96,Norman98,Chatterjee06,Shi08,Sassa11,Kaminski15,Loret17,Loret18,He14,Comin14,Comin16}.
However, the most of the spectral weight on the Fermi arcs locates at around the
tips of the Fermi arcs, indicating that the electrons at around the tips of the
Fermi arcs have a largest density of states
\cite{Chatterjee06,Shi08,Sassa11,Kaminski15,Loret17,Loret18,He14,Comin14,Comin16},
and then charge order is driven by this EFS instability, with a characteristic wave
vector corresponding to the tips of the Fermi arcs\cite{Feng16,Comin14,Comin16}.

The elementary excitations are parameterized by the electron spectral function
(\ref{ESF}), which has given a consistent description of the renormalization of
the electrons in the underdoped pseudogap phase\cite{Feng16,Liu21,Cao21,Zeng22},
where the obtained results show that the single-electron scattering rate has a
well-pronounced peak structure at around the antinodal and nodal regions, which
leads to the remarkable peak-dip-hump structure in the energy distribution curve.
Moreover, the dispersion kink is induced by the inflection point in the
single-electron scattering rate, while the spectral weight at around the
dispersion kink is reduced highly by the corresponding peak in the real part of the
electron normal self-energy. All these results are in qualitative agreement with the
corresponding experimental results
\cite{Dessau91,Randeria95,Fedorov99,Campuzano99,DMou17,Bogdanov00,Kaminski01,Johnson01,Iwasawa08,Plumb13}.
More importantly, the anisotropic suppression of the electronic density of states
on EFS by the normal-state pseudogap can affect the electrical transport in two ways
\cite{Timusk99}: (i) through the reduction of the number of electron current-carrying
states as a normal-state pseudogap forms; and (ii) since the electron current
carriers are scattered by the spin excitation, through the reduction in the density
of spin excitations.

\subsection{Electrical transport due to umklapp scattering from a spin excitation}
\label{Boltzmann-theory}

We now turn to discuss the low-temperature electrical transport in the underdoped
pseudogap phase. Theoretically, the Boltzmann transport equation is the cornerstone
for the discussions of the electrical transport\cite{Abrikosov88,Mahan81}, since
the Boltzmann transport equation is valid in the case of either the existence of the
well-defined quasiparticles or the treatment of the electron interaction mediated by
different bosonic modes within the Eliashberg approach. This builds on the following
pioneering works: (i) in the early days of the electrical transport research in the
conventional superconductors\cite{Prange64}, Prange and Kadanoff demonstrated that
in an electron-phonon system, a set of transport equations can be derived in the
Migdal's approximation, where the electron interaction mediated by the phonon leads
to the electron self-energy and vertex correction. In particular, this coupled set
of transport equations for the electron and phonon distribution functions is correct
even in the case of the absence of the well-defined quasiparticles\cite{Prange64}.
Nevertheless, one of the forms of the electrical transport equation,
\begin{eqnarray}\label{Boltzmann-equation-2}
e{\bf E}\cdot\nabla_{\bf k}f({\bf k})=I_{\rm e-e},
\end{eqnarray}
is identical to the electrical Boltzmann transport equation suggested by Landau for
the case in which the quasiparticle is well-defined\cite{Abrikosov88,Mahan81}, with
the charge of an electron $e$, the electron distribution function in a homogeneous
system $f({\bf k},t)$, and the electron-electron collision term $I_{\rm e-e}$. For a
convenience in the following discussions, the external magnetic field ${\bf H}$ has
been dropped, and only an external electric field ${\bf E}$ is applied to the system;
(ii) however, this Boltzmann transport equation (\ref{Boltzmann-equation-2}) is not
specific to the electron interaction mediated by the phonon in the conventional
superconductors, and has been confirmed recently that it is also valid for the
system with the electron interaction mediated by other bosonic excitations
\cite{Lee21,Ma23}. With the advantage of their insight\cite{Lee21,Ma23,Prange64},
the electrical transport due to the electron scattering mediated by various kinds
of bosonic modes can be evaluated in a simple way even in the case of the break down
of the quasiparticle picture.

\begin{figure}[h!]
\centering
\includegraphics[scale=0.80]{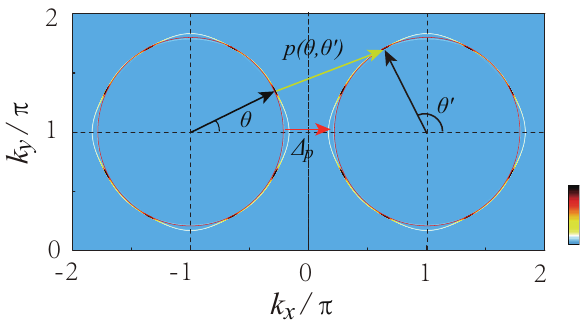}
\caption{(Color online) Schematic picture of the electron umklapp scattering process,
where an electron on a electron Fermi surface (left) is scattered by its partner on
the umklapp electron Fermi surface (right). The Fermi wave vector of the tips of the
Fermi arcs ${\rm k}^{\rm TFA}_{\rm F}$ is the radius
of the circular electron Fermi surface (red), and then an electron on this circular
electron Fermi surface (left) parametrized by the Fermi angle $\theta$ is scattered
to a point parametrized by the Fermi angle $\theta'$ on the umklapp electron Fermi
surface (right) by the spin excitation carrying momentum ${\rm p}(\theta,\theta')$.
$\Delta_{p}$ is the minimal umklapp vector at the antinode (the Fermi angle
$\theta=0$).\label{scatter-process}}
\end{figure}

In the following discussions, we study the the low-temperature resistivity in the
underdoped pseudogap phase based on the Boltzmann transport equation
(\ref{Boltzmann-equation-2}). For the calculation of this Boltzmann transport
equation (\ref{Boltzmann-equation-2}), the linear perturbation from the
equilibrium in terms of the distribution function has been introduced
\cite{Lee21,Ma23,Prange64},
\begin{eqnarray}\label{distribution-function}
f({\bf k})&=&n_{\rm F}({\bar{\varepsilon}_{\bf k}})
-{d n_{\rm F}({\bar{\varepsilon}_{\bf k}})
\over d{\bar{\varepsilon}_{\bf k}}}\tilde{\Phi}({\bf k}),
\end{eqnarray}
with the fermion distribution functions $n_{\rm F}(\omega)$, and the local shift of
the chemical potential at a given patch of EFS $\tilde{\Phi}({\bf k})$, which
satisfies the antisymmetric relation $\tilde{\Phi}(-{\bf k})=-\tilde{\Phi}({\bf k})$.
We substitute the above result in Eq. (\ref{distribution-function}) into
Eq. (\ref{Boltzmann-equation-2}), and then linearize the Boltzmann
equation (\ref{Boltzmann-equation-2}) as,
\begin{eqnarray}\label{Boltzmann-equation-3}
e{\bf v}_{\bf k}\cdot{\bf E}{dn_{\rm F}({\bar{\varepsilon}_{\bf k}})\over
d{\bar{\varepsilon}_{\bf k}}}=I_{\rm e-e},
\end{eqnarray}
where ${\bf v}_{\bf k}=\nabla_{\bf k}{\bar{\varepsilon}_{\bf k}}$ is the electron
velocity.

The electron-electron collision term $I_{\rm e-e}$ in the Boltzmann equation is
directly connected with the electrical scattering mechanism
\cite{Abrikosov88,Mahan81}. Although there is no consensus on the origin of the
electrical transport in the underdoped regime to date, it is widely believed that
the opening of the spin pseudogap induces a deviation from the T-linear behaviour
of the resistivity\cite{Gurvitch87,Bucher93,Ito93,Nakano94,Ando01}. In particular,
it has been shown clearly that the electron umklapp scattering is the origin of
the low-temperature T-linear resistivity in the strange-metal phase
\cite{Lee21,Honerkamp01,Hartnoll12}. Moreover, we\cite{Ma23} have also shown very
recently that the low-temperature T-linear resistivity in the overdoped
strange-metal phase originates from the umklapp scattering between electrons by the
exchange of the MF effective spin propagator. In the following discussions, we will
show that when this electron umklapp scattering mediated by the spin excitation in
the overdoped regime\cite{Ma23} flows to the underdoped regime, the opening of the
momentum dependence of the spin pseudogap naturally leads to a low-temperature
T-quadratic behaviour of the resistivity. To see the electron umklapp scattering
process more clearly, we
show a schematic picture of the electron umklapp scattering process\cite{Lee21,Ma23}
in Fig. \ref{scatter-process}, where an electron on a circular EFS (left) is
scattered by its partner on the umklapp EFS (right). It should be noted that the
intensity map of EFS in Fig. \ref{scatter-process} is the same as the EFS map in
Fig. \ref{EFS-map}, while the Fermi wave vector of the tips of the Fermi arcs
${\rm k}^{\rm TFA}_{\rm F}$ is the radius of the EFS circle (red). Moreover,
this circle EFS (red) connects all tips of the Fermi arcs, and then the most of the
electronic density of states is concentred on this circular EFS.

Following these recent discussions\cite{Lee21,Ma23}, we can derive the
electron-electron collision $I_{\rm e-e}$ in Eq. (\ref{Boltzmann-equation-3}) in the
underdoped pseudogap phase as,
\begin{widetext}
\begin{eqnarray}\label{electron-collision-1}
I_{\rm e-e}&=&{1\over N^{2}}\sum_{{\bf k}',{\bf p}} {2\over T}
|P({\bf k},{\bf p},{\bf k}',\bar{\varepsilon}_{\bf k}
-\bar{\varepsilon}_{{\bf k}+{\bf p}+{\bf G}})|^{2}
\{\tilde{\Phi}({\bf k})+\tilde{\Phi}({\bf k'})
-\tilde{\Phi}({\bf k}+{\bf p}+{\bf G})-\tilde{\Phi}({\bf k}'-{\bf p})\}\nonumber\\
&\times& n_{\rm F}(\bar{\varepsilon}_{\bf k})
n_{\rm F}(\bar{\varepsilon}_{{\bf k}'})
[1-n_{\rm F}(\bar{\varepsilon}_{{\bf k}+{\bf p}+{\bf G}})]
[1-n_{\rm F}(\bar{\varepsilon}_{{\bf k}'-{\bf p}})]
\delta(\bar{\varepsilon}_{\bf k}+\bar{\varepsilon}_{\bf k'}
-\bar{\varepsilon}_{{\bf k}+{\bf p}+{\bf G}}-\bar{\varepsilon}_{{\bf k}'-{\bf p}}),
\end{eqnarray}
\end{widetext}
where ${\bf G}$ represents a set of reciprocal lattice vectors. It should be
emphasized that the above electron umklapp scattering (\ref{electron-collision-1})
is described as a scattering between electrons by the exchange of the effective
spin propagator,
\begin{eqnarray}\label{ESP}
P({\bf k},{\bf p},{\bf k}',\omega)&=&{1\over N}\sum_{\bf q}
\Lambda_{{\bf p}+{\bf q}+{\bf k}}\Lambda_{{\bf q}+{\bf k}'}
\bar{\Pi}({\bf p},{\bf q},\omega),~~~
\end{eqnarray}
rather than the scattering between electrons via the emission and absorption of the
spin excitation\cite{Lee21,Ma23}, where
$\Lambda_{{\bf k}}=4t\gamma_{\bf k}-4t'\gamma_{\bf k}'$ is the bare vertex function.

In cuprate superconductors, a small density of charge carriers is sufficient to
destroy the AF long-range order (AFLRO). However, the doped charge carriers and
the coupled spins organize themselves in a cooperative way to enhance
both the electrons mobility and the AF short-range order (AFSRO) correlation, and
then the spin excitations in the spin liquid state with AFSRO appear to
survive from the underdoped regime to the overdoped regime\cite{Fujita12}. Moreover,
it has been shown that the coupling strength of the electrons with the spin
excitations gradually weakens with the increase of doping from a strong-coupling
case in the underdoped regime to a weak-coupling side in the overdoped regime
\cite{Feng12,Johnson01,Kordyuk10}, reflecting a reduction of the strength of the
magnetic fluctuation with the increase of doping. In other words, (i) the effect of
the magnetic fluctuation in the overdoped regime is less dramatic than in the
underdoped regime\cite{Kastner98,Timusk99,Hufner08,Vishik18,Fujita12}; (ii) on the
other hand, as shown in Fig. \ref{spin-pseudogap-doping}, the effect from antinodal
spin pseudogap is absent in the overdoped regime, and then the main properties of
the antinodal spin excitations can be well described by the MF spin propagator.
These are reasons why the interpretation of the low-temperature
T-linear resistivity in the overdoped strange-metal phase can be well made in terms
of the umklapp scattering between electrons by the exchange of the MF effective
spin propagator\cite{Ma23}.

However, in the strong-coupling side (then in the underdoped regime), (i) the effect
of the magnetic fluctuation is much dramatic
\cite{Kastner98,Timusk99,Hufner08,Vishik18,Fujita12}; and (ii) as shown in
Fig. \ref{spin-pseudogap-doping}, the effect from antinodal spin pseudogap is
particularly notable, where the antinodal spin pseudogap lowers the density of the
spin excitations\cite{Kuang15} in response to the intense electron umklapp
scattering, which reduces the strength of the electron umklapp scattering from the
spin excitation states into the antinodal region. In this case, the umklapp
scattering between electrons should be mediated by the exchange of the full
effective spin propagator for a proper description of the low-temperature
resistivity in the underdoped pseudogap phase.

Now our aim is to obtain the full effective spin propagator. The full spin bubble
$\bar{\Pi}({\bf p},{\bf q},\omega)$ in Eq. (\ref{ESP}) is a convolution of
two full spin propagators, and can be derived directly from the full spin
propagator (\ref{FSGF-spin-gap}) as,
\begin{equation}\label{spin-bubble}
\bar{\Pi}({\bf p},{\bf q},\omega)=-\sum_{\substack{\alpha=1,2\\ \alpha'=1,2}}
{\bar{W}^{(1)}_{\alpha\alpha'{\bf p}{\bf q}}\over\omega^{2}
-[\bar{\omega}^{(1)}_{\alpha\alpha'{\bf p}{\bf q}}]^{2}}
+{\bar{W}^{(2)}_{\alpha\alpha'{\bf p}{\bf q}}\over\omega^{2}
-[\bar{\omega}^{(2)}_{\alpha\alpha'{\bf p}{\bf q}}]^{2}},~~~~~~
\end{equation}
where the spin excitation energy dispersions
$\bar{\omega}^{(1)}_{\alpha\alpha'{\bf p}{\bf q}}$ and
$\bar{\omega}^{(2)}_{\alpha\alpha'{\bf p}{\bf q}}$ are given by,
\begin{subequations}\label{Effective-FSEED}
\begin{eqnarray}
\bar{\omega}^{(1)}_{\alpha\alpha'{\bf p}{\bf q}}
&=&\bar{\omega}_{\alpha{\bf q}+{\bf p}}+\bar{\omega}_{\alpha'{\bf q}},~~~~\\
\bar{\omega}^{(2)}_{\alpha\alpha'{\bf p}{\bf q}}
&=&\bar{\omega}_{\alpha{\bf q}+{\bf p}}-\bar{\omega}_{\alpha'{\bf q}},
\end{eqnarray}
\end{subequations}
respectively, and the corresponding functions,
\begin{eqnarray}
\bar{W}^{(1)}_{\alpha\alpha'{\bf p}{\bf q}}&=&{\bar{B}_{\alpha'{\bf q}}
\bar{B}_{\alpha{\bf q}+{\bf p}}\over 2\bar{\omega}_{\alpha'{\bf q}}
\bar{\omega}_{\alpha{\bf q}+{\bf p}}}\bar{\omega}^{(1)}_{\alpha\alpha'{\bf p}{\bf q}}
\nonumber\\
&\times& [n_{\rm B}(\bar{\omega}_{\alpha{\bf q}+{\bf p}})
+n_{\rm B}(\bar{\omega}_{\alpha'{\bf q}})+1], ~~~~~\\
\bar{W}^{(2)}_{\alpha\alpha'{\bf p}{\bf q}}&=&{\bar{B}_{\alpha'{\bf q}}
\bar{B}_{\alpha{\bf q}+{\bf p}}\over 2\bar{\omega}_{\alpha'{\bf q}}
\bar{\omega}_{\alpha{\bf q}+{\bf p}}}\bar{\omega}^{(2)}_{\alpha\alpha'{\bf p}{\bf q}}
\nonumber\\
&\times& [n_{\rm B}(\bar{\omega}_{\alpha{\bf q}+{\bf p}})
-n_{\rm B}(\bar{\omega}_{\alpha'{\bf q}})],
\end{eqnarray}
where $n_{\rm B}(\omega)$ is the boson distribution function.
From the above spin bubble (\ref{spin-bubble}), the full effective spin propagator
(\ref{ESP}) now can be derived as,
\begin{eqnarray}\label{reduced-propagator}
P({\bf k},{\bf p},{\bf k}',\omega)&=&-{1\over N}\sum\limits_{\alpha\alpha'{\bf q}}
\left [ {\varpi^{(1)}_{\alpha\alpha'}({\bf k},{\bf p},{\bf k}',{\bf q})\over
\omega^{2}-[\bar{\omega}^{(1)}_{\alpha\alpha'{\bf p}{\bf q}}]^{2}}\right .
\nonumber\\
&-& \left . {\varpi^{(2)}_{\alpha\alpha'}({\bf k},{\bf p},{\bf k}',{\bf q})\over
\omega^{2}-[\bar{\omega}^{(2)}_{\alpha\alpha'{\bf p}{\bf q}}]^{2}} \right ],~~~~~
\end{eqnarray}
with the weight functions,
\begin{subequations}
\begin{eqnarray}
\varpi^{(1)}_{\alpha\alpha'}({\bf k},{\bf p},{\bf k}',{\bf q})&=&
\Lambda_{{\bf k}+{\bf p}+{\bf q}}
\Lambda_{{\bf q}+{\bf k}'}\bar{W}^{(1)}_{\alpha\alpha'{\bf p}{\bf q}},\\
\varpi^{(2)}_{\alpha\alpha'}({\bf k},{\bf p},{\bf k}',{\bf q})&=&
\Lambda_{{\bf k}+{\bf p}+{\bf q}}
\Lambda_{{\bf q}+{\bf k}'}\bar{W}^{(2)}_{\alpha\alpha'{\bf p}{\bf q}}.
\end{eqnarray}
\end{subequations}

The electron umklapp scattering in Eq. (\ref{electron-collision-1}) shows that the
electron-electron collision $I_{\rm e-e}$ is both functions of momentum and energy.
Howerver, at low temperatures, everything happens near EFS\cite{Abrikosov88,Mahan81}.
In this case, an any given patch on the circular EFS shown in
Fig. \ref{scatter-process} is represented via the Fermi angle $\theta$ with the Fermi
angle range $\theta\in [0,2\pi]$, and then the momentum integration along the
perpendicular momentum can be replaced by the integration\cite{Lee21,Ma23,Prange64}
over $\bar{\varepsilon}_{\bf k}$. For the umklapp scattering between electrons by
the exchange of the full effective spin propagator in
Eq. (\ref{electron-collision-1}), an electron on the circular EFS parametrized by
the Fermi angle $\theta$ is scattered to a point parametrized by the Fermi angle
$\theta'$ on the umklapp EFS via the spin excitation carrying momentum
${\rm p}(\theta,\theta')$ as shown in Fig. \ref{scatter-process}. Following the
above treatment and the calculation process in the recent works\cite{Lee21,Ma23}
for the low-temperature resistivity in the overdoped strange-metal phase, the
electron-electron collision $I_{\rm e-e}$ in Eq. (\ref{electron-collision-1}) in
the underdoped pseudogap phase can be obtained straightforwardly\cite{Ma23}, and
then the Boltzmann transport equation (\ref{Boltzmann-equation-3}) can be derived
as,
\begin{eqnarray}\label{electron-collision}
e{\bf v}_{\rm F}(\theta)\cdot {\bf E}=-2\int {d\theta'\over {2\pi}}\zeta(\theta')
F(\theta,\theta')[\Phi(\theta)-\Phi(\theta')],~~~~~
\end{eqnarray}
with $\Phi(\theta)=\tilde{\Phi}[{\rm k}(\theta)]$, the Fermi velocity
${\bf v}_{\rm F}(\theta)$ at the Fermi angle $\theta$, the density of states factor
$\zeta(\theta')={\rm k}^{2}_{\rm F}/[4\pi^{2}{\rm v}^{3}_{\rm F}]$ at angle
$\theta'$, the Fermi wave vector ${\rm k}_{\rm F}$, and the Fermi velocity
${\rm v}_{\rm F}$. In particular, the antisymmetric relation
$\tilde{\Phi}(-{\bf k})=-\tilde{\Phi}({\bf k})$ for $\tilde{\Phi}({\bf k})$ in
Eq. (\ref{distribution-function}) has been converted into
$\Phi(\theta)=-\Phi(\theta+\pi)$ for $\Phi(\theta)$ in the above
Eq. (\ref{electron-collision}). Moreover, the coefficient of $\Phi(\theta)$ in the
first term of the right-hand side of Eq. (\ref{electron-collision}),
\begin{eqnarray}\label{scattering-rate}
\gamma(\theta)=2\int {d \theta' \over {2\pi}} \zeta(\theta')F(\theta,\theta'),
\end{eqnarray}
can be referred to as the angular (momentum) dependence of the umklapp scattering
rate\cite{Ma23,Lee21}, with the kernel function $F(\theta,\theta')$ that connects
the point $\theta$ on the circular EFS with the point $\theta'$ on the umklapp EFS
via the amplitude of the momentum transfer ${\rm p}(\theta,\theta')$ as shown in
Fig. \ref{scatter-process}, which can be derived as,
\begin{eqnarray}\label{kernel-function}
F(\theta,\theta')&=&{1\over T}\int {d\omega\over 2\pi}{\omega^{2}\over
{\rm p}(\theta,\theta')}
{|\bar{P}[{\rm k}(\theta),{\rm p}(\theta,\theta'),\omega]|}^{2}\nonumber\\
&\times& n_{\rm B}(\omega)[1+n_{\rm B}(\omega)],~~~~~~
\end{eqnarray}
where the full effective spin propagator
$\bar{P}[{\rm k}(\theta),{\rm p}(\theta,\theta'),\omega]$ has been rewritten
explicitly in terms of the Fermi angles $\theta$ and $\theta'$ as\cite{Ma23},
\begin{eqnarray}\label{reduced-propagator-5}
\bar{P}[{\rm k}(\theta),{\rm p}(\theta,\theta',\omega]&=&-{1\over N}
\sum\limits_{\alpha\alpha'{\bf q}}\left [
{\varpi^{(1)}_{\alpha\alpha'}(\theta,\theta',{\bf q})\over\omega^{2}
-[\bar{\omega}^{(1)}_{\alpha\alpha'\theta,\theta'}({\bf q})]^{2}}\right .\nonumber\\
&-& \left . {\varpi^{(2)}_{\alpha\alpha'}(\theta,\theta',{\bf q})\over\omega^{2}
-[\bar{\omega}^{(2)}_{\alpha\alpha'\theta,\theta'}({\bf q})]^{2}} \right ],~~~~~
\end{eqnarray}
where $\varpi^{(1)}_{\alpha\alpha'}(\theta,\theta',{\bf q})=
\varpi^{(1)}_{\alpha\alpha'}
[{\rm k}(\theta),{\rm p}(\theta,\theta'),{\bf k}'_{\rm F},{\bf q}]$,
$\varpi^{(2)}_{\alpha\alpha'}(\theta,\theta',{\bf q})=
\varpi^{(2)}_{\alpha\alpha'}
[{\rm k}(\theta),{\rm p}(\theta,\theta'),{\bf k}'_{\rm F},{\bf q}]$,
$\bar{\omega}^{(1)}_{\alpha\alpha'\theta,\theta'}({\bf q})
=\bar{\omega}^{(1)}_{\alpha\alpha'{\rm p}(\theta,\theta'){\bf q}}$, and
$\bar{\omega}^{(2)}_{\alpha\alpha'\theta,\theta'}({\bf q})
=\bar{\omega}^{(2)}_{\alpha\alpha'{\rm p}(\theta,\theta'){\bf q}}$.

At low temperatures, the electron elastic scattering occurs on EFS, while the
electron inelastic scattering occurs near EFS\cite{Kohsaka08,Abrikosov88,Mahan81}.
In this case, the above energy integration in the kernel function
(\ref{kernel-function}) actually includes the inelastic scattering process of
momentum and energy exchange
between electrons\cite{Lee21,Ma23,Prange64}. This follows from a fact that the
umklapp scattering rate $\gamma(\theta)$ in Eq. (\ref{scattering-rate}) is
directly associated with the kernel function $F(\theta,\theta')$ in
Eq. (\ref{kernel-function}), while this kernel function is obtained by the
integration of the spin excitation energy $\omega$, leading to the occurrence of
the inelastic scattering process near EFS in the present umklapp
scattering. In particular, in the case of the absence of the spin pseudogap, this
electron umklapp scattering induces a low-temperature T-linear resistivity in the
overdoped regime\cite{Ma23}, in agreement with the experimental observations
\cite{Legros19,Ayres21,Grisso21}, where the experimental analyses seem to indicate
that the T-linear resistivity is due to the inelastic scattering.

For the discussions of the low-temperature resistivity, we need to obtain the
electron current density, which can be derived in terms of the local shift of the
chemical potential $\Phi(\theta)$ as\cite{Ma23},
\begin{eqnarray}\label{current-density}
{\bf J} &=& en_{0}{1\over N}\sum_{\bf k}{\bf v}_{\bf k}
{dn_{\rm F}({\bar{\varepsilon}_{\bf k}})\over d\bar{\varepsilon}_{\bf k}}
\tilde{\Phi}({\bf k})\nonumber\\
&=& -en_{0}{{\rm k}_{\rm F}\over {\rm v}_{\rm F}}\int
{d\theta\over (2\pi)^{2}}{\bf v}_{\rm F}(\theta)\Phi(\theta),~~~~~
\end{eqnarray}
with the momentum relaxation that is generated by the action of the electric field
on the mobile electrons at EFS with the density $n_{0}$. However, this local shift
of the chemical potential $\Phi(\theta)$ can be evaluated directly in the
relaxation-time approximation as\cite{Lee21,Ma23},
$\Phi(\theta)=-e{\rm v}_{\rm F}{\rm cos}(\theta)E_{\hat{x}}/[2\gamma(\theta)]$,
where the electric field ${\bf E}$ has been selected along the $\hat{x}$-axis.
With the help of the above treatment, the dc conductivity can be derived as
\cite{Lee21,Ma23},
\begin{eqnarray}\label{dc-conductivity}
\sigma_{\rm dc}(T)={1\over 2}e^{2}n_{0}{\rm k}_{\rm F}{\rm v}_{\rm F}
\int {d\theta\over (2\pi)^{2}}{\rm cos}^{2}(\theta)
{1\over\gamma(\theta)},
\end{eqnarray}
and then the resistivity can be expressed directly in terms of the above
dc conductivity as,
\begin{eqnarray}\label{dc-resistivity}
\rho(T)={1\over \sigma_{\rm dc}(T)}.
\end{eqnarray}
It thus shows that the electrical resistance originates from the umklapp scattering
between electrons by the exchange of the full effective spin propagator.

\section{Quantitative characteristics}\label{electron-resistivity}

\begin{figure}[h!]
\centering
\includegraphics[scale=0.69]{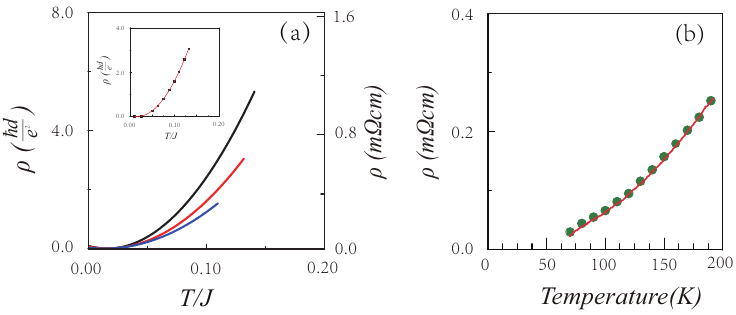}
\caption{(Color online) (a) Resistivity as a function of temperature at
$\delta=0.06$ (black-line), $\delta=0.09$ (red-line), and $\delta=0.12$ (blue-line).
Inset: the numerical fit of the resistivity (black-dots) at $\delta=0.09$ with the
fit form $\rho(T)=A_{2}T^{2}$. (b) The corresponding experimental result of the
low-temperature T-quadratic resistivity for the underdoped
HgBa$_{2}$CuO$_{4+\delta}$ taken from Ref. \onlinecite{Mirzaei13}.
\label{resistivity-temperature}}
\end{figure}

It should be emphasized that the above resistivity in Eq. (\ref{dc-resistivity})
is obtained in the pure two-dimensional $t$-$J$ model (\ref{tJ-model}) on a
square lattice. However, for a clear comparison with the corresponding
experimental data
\cite{Ando04a,Ando04,Lee05,Proust08,Cooper09,Hussey11,Mirzaei13,Barisic13,Pelc20},
the resistivity obtained in the above equation (\ref{dc-resistivity}) should be
renormalized by the distance between the adjacent copper-oxide layers as it has
been done in Ref. \onlinecite{Bergeron11}, with the interlayer lattice constant
\cite{Bergeron11} that can be chosen as $d=0.5$nm. Now we are ready to discuss the
low-temperature resistivity in the underdoped pseudogap phase. In
Fig. \ref{resistivity-temperature}a, we plot the renormalized resistivity $\rho(T)$
as a function of temperature at doping $\delta=0.06$ (black-line), $\delta=0.09$
(red-line), and $\delta=0.12$ (blue-line), where in the low-temperature region, a
power-law resistivity appears over a wide doping range in the underdoped regime.
To explore this power-law behaviour of the low-temperature resistivity more clearly,
the above results of $\rho(T)$ in Fig. \ref{resistivity-temperature}a have been
numerically fitted with the fit form $\rho(T)=A_{2}T^{2}$, and the fitted result of
the low-temperature resistivity (black-dots) at $\delta=0.09$ is also plotted in
Fig. \ref{resistivity-temperature}a (inset), where the low-temperature resistivity
is demonstrated clearly to be grown quadratically as the temperature is raised.
For a better comparison, the corresponding experimental result\cite{Mirzaei13} of
the low-temperature T-quadratic resistivity observed in the underdoped
HgBa$_{2}$CuO$_{4+\delta}$ is also shown in Fig. \ref{resistivity-temperature}b,
where the most characteristic feature of the low-temperature resistivity in the
underdoped regime is that it is perfectly quadratic down to the lowest achievable
temperatures
\cite{Ando04a,Ando04,Lee05,Proust08,Cooper09,Hussey11,Mirzaei13,Barisic13,Pelc20}.
Apparently, (i) {\it this characteristic feature of the T-quadratic behaviour of
the low-temperature resistivity is the same in the theory and experiments}
\cite{Ando04a,Ando04,Lee05,Proust08,Cooper09,Hussey11,Mirzaei13,Barisic13,Pelc20};
(ii) the magnitude of the low-temperature T-quadratic resistivity at a given doping
and a given temperature is also qualitatively consistent with the corresponding
experimental results in the underdoped regime
\cite{Proust08,Cooper09,Hussey11,Mirzaei13,Barisic13,Pelc20}, where
different magnitudes at a given doping and a given temperature have been observed
for different families of cuprate superconductors.
\begin{figure}[h!]
\centering
\includegraphics[scale=0.725]{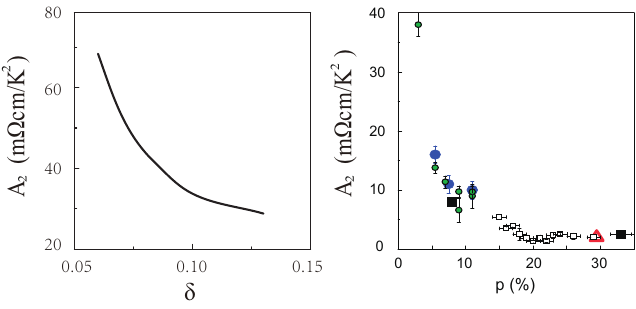}
\caption{(a) Strength of the T-quadratic resistivity as a function of doping. (b)
The corresponding experimental result of cuprate superconductors taken from
Ref. \onlinecite{Barisic13}.\label{coefficient-doping}}
\end{figure}
Moreover, the strength of the T-quadratic resistivity (then the T-quadratic
resistivity coefficient) $A_{2}$ grows as the doping is reduced.
To see this doping
dependence of the T-quadratic resistivity strength more clearly, we plot $A_{2}$ as
a function of doping $\delta$ in Fig. \ref{coefficient-doping}a in comparison with
the corresponding experimental results\cite{Barisic13} of the underdoped cuprate
superconductors in Fig. \ref{coefficient-doping}b. It thus shows clearly
that $A_{2}$ {\it drops} gradually as doping is enhanced in the underdoped regime,
indicating that $A_{2}$ is roughly proportional to the inverse of the doping
concentration. This {\it tendency of the doping dependent} $A_{2}$ is also in
{\it qualitative} agreement with the corresponding experimental results of cuprate
superconductors in the underdoped regime
\cite{Ando04a,Ando04,Lee05,Proust08,Cooper09,Hussey11,Mirzaei13,Barisic13,Pelc20}.
These results in Fig. \ref{resistivity-temperature} and
Fig. \ref{coefficient-doping} in the underdoped regime along with the recent
results\cite{Ma23} of the low-temperature T-linear resistivity in the overdoped
regime therefore show that the electron umklapp scattering from a spin excitation
responsible for the low-temperature T-linear resistivity in the overdoped regime
naturally produces the low-temperature T-quadratic resistivity in the underdoped
regime due to the opening of the momentum dependence of the spin pseudogap.

The low-temperature resistivity is mainly determined by the umklapp scattering rate
$\gamma(\theta,T)$ in Eq. (\ref{scattering-rate}), however, this umklapp scattering
rate $\gamma(\theta,T)$ is in turn directly connected with the kernel function
$F(\theta,\theta')$ (then the probability weight of the umklapp scattering) in
Eq. (\ref{kernel-function}) as we have mentioned above. For the further understanding
of the nature of the umklapp scattering between electrons by the exchange of the full
effective spin propagator, we first analyse the exotic feature of the kernel function
$F(\theta,\theta')$.
\begin{figure}
\centering
\includegraphics[scale=0.95]{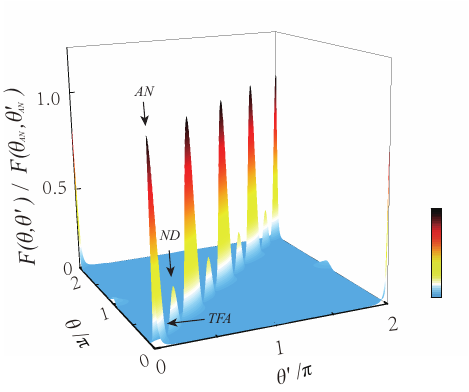}
\caption{(Color online) Surface plot of the kernel function
$F(\theta,\theta')/F(\theta_{\rm AN},\theta'_{\rm AN})$ at
$\delta=0.09$ with $T=0.05J$, where AN, TFA, and ND denote the antinode, tip of the
Fermi arc, and node, respectively, while $F(\theta_{\rm AN},\theta'_{\rm AN})$ is
the magnitude of $F(\theta,\theta')$ at the antinode. \label{kernel}}
\end{figure}
\begin{figure}[h!]
\centering
\includegraphics[scale=0.95]{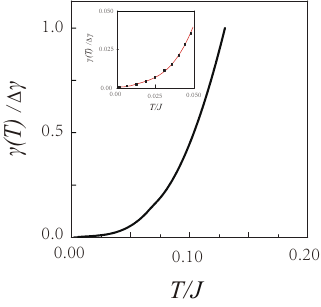}
\caption{Umklapp scattering rate $\gamma(T)/\Delta\gamma$ at the antinode as a
function of temperature for $\delta=0.09$, where
$\Delta\gamma=\gamma(T_{0})-\gamma(0)$, with $T_{0}=0.13J$. The inset shows the
detail of the temperature dependence of $\gamma(T)$ at the antinode in the far
lower temperature region. \label{out-rate-temperature}}
\end{figure}
In Fig. \ref{kernel}, we plot $F(\theta,\theta')$ at
$\delta=0.09$ with $T=0.05J$, where the probability weight of the electron umklapp
scattering is very strongly anisotropic in momentum space.
In particular, as in the
case of the overdoped strange-metal phase\cite{Ma23}, the almost all of the
probability weight of the electron umklapp scattering in the underdoped pseudogap
phase is also concentrated at around the antinodal region, leading to that the
strongest umklapp scattering occurs at around this antinodal region. However, the
weakest umklapp scattering appears at around the tips of the Fermi arcs, since the
probability weight of the electron umklapp scattering there is almost none.
Moreover, the strength of the electron umklapp scattering at around the nodal region
is much weaker than that at around the antinodal region, since a very small amount
of the probability weight of the electron umklapp scattering is concentrated at
around the nodal region. This special momentum dependent distribution of the
probability weight of the electron umklapp scattering in Fig. \ref{kernel} together
with the recent discussions of the momentum-dependent distribution of the probability
weight of the electron umklapp scattering in the overdoped regime\cite{Ma23}
therefore confirm that in the whole doping regime, the dominant contribution to the
scattering rate mainly arises from the umklapp scattering between the relatively slow
electrons at around the antinodal region.

We now turn to explore the unusual evolution of the umklapp scattering rate
(\ref{scattering-rate}) with temperature in the underdoped pseudogap phase. We have
made a series of calculations for the umklapp scattering rate $\gamma(T)$ at
different Fermi angles, and the result of $\gamma(T)$ as a function of temperature
for $\delta=0.09$ at the antinode is plotted in Fig. \ref{out-rate-temperature},
where the inset shows the detail of the evolution of $\gamma(T)$ with temperature at
the antinode in the far lower temperature region. Our numerical fit demonstrates that
$\gamma(T)$ is purely T-quadratic in the
low-temperature region, i.e., it increases quadratically with temperature as the
temperature is increased. In particular, as the case of the resistivity shown in the
inset of Fig. \ref{resistivity-temperature}a, this $\gamma(T)$ approaches zero
as the temperature approaches zero. Moreover, although the magnitude of
$\gamma(\theta,T)$ is strongly anisotropic in momentum space, the
low-temperature T-quadratic behaviour of $\gamma(\theta,T)$ appears at an any given
Fermi angle $\theta$. Comparing this result in Fig. \ref{out-rate-temperature} with
the corresponding results of the low-temperature T-quadratic resistivity in
Fig. \ref{resistivity-temperature}a, it thus shows that the low-temperature
T-quadratic behaviour of $\gamma(T)$ together with the temperature region are the
same as the corresponding behaviour and region in the resistivity $\rho(T)$, which
therefore confirms that the low-temperature T-quadratic resistivity with the
corresponding temperature region is mainly governed by the low-temperature
T-quadratic umklapp scattering rate with the corresponding temperature region.

Now we give an explanation to show why the low-temperature resistivity has a
T-quadratic behaviour in the underdoped pseudogap phase, with a dramatic switch to
the T-linear behaviour in the overdoped strange-metal phase? The nature of the
umklapp scattering rate in Eq. (\ref{scattering-rate}) is mainly determined by the
nature of the kernel function $F(\theta,\theta')$ in Eq. (\ref{kernel-function}),
however, this kernel function $F(\theta,\theta')$ in the underdoped pseudogap phase
is proportional to the full effective spin propagator
$P({\bf k},{\bf p},{\bf k}',\omega)$ in Eq. (\ref{ESP}). From the full spin
propagator in Eq. (\ref{FSGF-spin-gap}), the full spin spectral function is derived
straightforwardly as $A_{\rm spin}({\bf k},\omega)=A^{(1)}_{\rm spin}({\bf k},\omega)
+A^{(2)}_{\rm spin}({\bf k},\omega)$, where the components
$A^{(1)}_{\rm spin}({\bf k},\omega)$ and $A^{(2)}_{\rm spin}({\bf k},\omega)$
are given by,
\begin{subequations}\label{full-spectral-function}
\begin{eqnarray}
A^{(1)}_{\rm spin}({\bf k},\omega)&=&{\bar{B}_{1{\bf k}}\over
\pi\bar{\omega}_{1{\bf k}}}
[\delta(\omega-\bar{\omega}_{1{\bf k}})-\delta(\omega+\bar{\omega}_{1{\bf k}})],
~~~~\label{full-spectral-function-1}\\
A^{(2)}_{\rm spin}({\bf k},\omega)&=& {\bar{B}_{2{\bf k}}\over
\pi\bar{\omega}_{2{\bf k}}}
[\delta(\omega-\bar{\omega}_{2{\bf k}})-\delta(\omega+\bar{\omega}_{2{\bf k}})]
\label{full-spectral-function-2},
\end{eqnarray}
\end{subequations}
respectively. However, during the calculation, we have found that the spectral
weight in the component $A^{(1)}_{\rm spin}({\bf k},\omega)$ is several orders of
magnitude greater than the corresponding spectral weight in the component
$A^{(2)}_{\rm spin}({\bf k},\omega)$, leading to that the electron umklapp
scattering is mainly mediated by the spin excitations from the component of
$A^{(1)}_{\rm spin}({\bf k},\omega)$.
\begin{figure}[h!]
\includegraphics[scale=0.90]{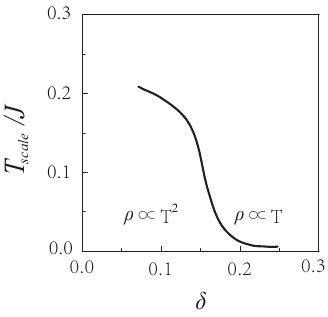}
\caption{The low-temperature scale $T_{\rm scale}$ as a function of doping.
\label{T-scale-doping}}
\end{figure}
On the other hand, the spin excitations with
the full spin excitation dispersion $\bar{\omega}_{1{\bf k}}$ around the AF wave
vector ${\bf k}_{\rm A}=[\pm\pi,\pm\pi]$ have the largest density of states. With
these special properties of the spin excitations in the underdoped pseudogap phase,
the full effective spin propagator $P({\bf k},{\bf p},{\bf k}',\omega)$ in
Eq. (\ref{reduced-propagator}) can be reduced approximately as,
\begin{eqnarray}\label{reduced-propagator-1}
P({\bf k},{\bf p},{\bf k}',\omega)&\approx&-{1\over N}\sum\limits_{\bf q}\left [
{\varpi^{(1)}_{11}({\bf k},{\bf p},{\bf k}',{\bf q})\over\omega^{2}
-[\omega^{(1)}_{11{\bf p}{\bf q}}]^{2}}\right .\nonumber\\
&-& \left . {\varpi^{(2)}_{11}({\bf k},{\bf p},{\bf k}',{\bf q})\over\omega^{2}
-[\omega^{(2)}_{11{\bf p}{\bf q}}]^{2}} \right ].~~~~~
\end{eqnarray}
Now we follow the discussions of the nature of the kernel function in
Ref. \onlinecite{Ma23} in the overdoped strange-metal phase to make a taylor
expansion for the spin excitation energy dispersions
$\bar{\omega}^{(1)}_{11{\bf p}{\bf q}}=\bar{\omega}_{1{\bf q}+{\bf p}}
+\bar{\omega}_{1{\bf q}}$ and $\bar{\omega}^{(2)}_{11{\bf p}{\bf q}}
=\bar{\omega}_{1{\bf q}+{\bf p}}-\bar{\omega}_{1{\bf q}}$ in
Eq. (\ref{Effective-FSEED}), and then the spin excitation energy dispersions
$\bar{\omega}^{(1)}_{11{\bf p}{\bf q}}$ and $\bar{\omega}^{(2)}_{11{\bf p}{\bf q}}$
can be expressed approximately as,
\begin{subequations}\label{effective-spin-excitation}
\begin{eqnarray}
\bar{\omega}^{(1)}_{11{\bf p}{\bf q}}&=&\bar{\omega}_{1{\bf q}+{\bf p}}
+\bar{\omega}_{1{\bf q}}\approx b_{\bf q}p^{2}+2\bar{\omega}_{1{\bf q}},\\
\bar{\omega}^{(2)}_{11{\bf p}{\bf q}}&=&\bar{\omega}_{1{\bf q}+{\bf p}}
-\bar{\omega}_{1{\bf q}}\approx b_{\bf q}p^{2},
\end{eqnarray}
\end{subequations}
with $b_{\bf q}=d^{2}\bar{\omega}_{1{\bf q}}/(d^{2}q)$. The above results in
Eq. (\ref{effective-spin-excitation}) show that as in the case of the the overdoped
strange-metal phase\cite{Ma23}, the full effective spin propagator
$P({\bf k},{\bf p},{\bf k}',\omega)$ in Eq. (\ref{reduced-propagator-1}) scales with
$p^{2}$, and then when the electron umklapp scattering kicks in, the energy scale is
proportional to $\Delta^{2}_{p}$ due to the presence of this $p^{2}$ scaling in the
full effective spin propagator (\ref{reduced-propagator-1}). In this case,
$T_{\rm scale}=\bar{b}\Delta^{2}_{p}$ can be identified as {\it the low-temperature
scale}, with the average value
$\bar{b}=(1/N)\sum\limits_{{\bf q}\in \{{\bf k}_{\rm A}\}}b({\bf q})$ that is a
constant at a given doping, where the summation ${\bf q}\in \{{\bf k}_{\rm A}\}$ is
restricted to the extremely small area $\{{\bf k}_{\rm A}\}$ around the
${\bf k}_{\rm A}$ point of BZ. However, this {\it low-temperature scale}
$T_{\rm scale}$ is strongly doping dependent. To see
this doping dependence of $T_{\rm scale}$ more clearly, we plot $T_{\rm scale}$ as a
function of doping in Fig. \ref{T-scale-doping}, where $T_{\rm scale}$ is relatively
high at around the slightly underdoped region, then it decreases when the doping is
increased in the heavily underdoped region, and is reduced to {\it a very low
temperature} in the overdoped regime. Moreover, $T_{\rm scale}$ {\it as a function
of doping} in Fig. \ref{T-scale-doping} {\it in the underdoped regime presents a
similar behavior of the antinodal spin pseudogap crossover temperature}
$T^{*}_{\rm s}$ shown in Fig. \ref{spin-pseudogap-doping}b, {\it suggesting a
possible correlation between} $T_{\rm scale}$ {\it and the opening of the antinodal
spin pseudogap below} $T^{*}_{\rm s}$ {\it in the underdoped pseudogap phase}.

With the help of the above full effective spin propagator
$P({\bf k},{\bf p},{\bf k}',\omega)$ in Eq. (\ref{reduced-propagator-1}) and the
doping dependence of $T_{\rm scale}$ in Fig. \ref{T-scale-doping}, we now follow
the similar analysis carried out in the overdoped strange-metal phase\cite{Ma23} to
show that two primary regions of the low-temperature resistivity need to be
distinguished:\\
(i) in the low-temperature T-quadratic region ($T<T_{\rm scale}$) in the underdoped
regime, the kernel function $F(\theta,\theta')$ is reduced as
$F(\theta,\theta')\propto T^{2}$, which naturally produces a low-temperature
T-quadratic resistivity $\rho(T)\propto T^{2}$ as shown in
Fig. \ref{resistivity-temperature}; \\
(ii) in the low-temperature T-linear region ($T>T_{\rm scale}$) in the overdoped
regime, the kernel function $F(\theta,\theta')$ is reduced as
$F(\theta,\theta')\propto T$, which naturally induces a T-linear resistivity
$\rho(T)\propto T$ as we have shown in Ref. \onlinecite{Ma23}. In particular, this
low-temperature scale $T_{\rm scale}$ in the overdoped regime is very low
\cite{Ma23} due to the absence of the antinodal spin pseudogap as shown in
Fig. \ref{spin-pseudogap-doping}.

\section{Summary and discussion}\label{summary}

Within the framework of the kinetic-energy-driven superconductivity, we have
rederived the full spin propagator, the full charge-carrier propagator, and the
full electron propagator in the normal-state of cuprate superconductors, where the
spin self-energy (then the spin pseudogap) is obtained explicitly in terms of the
collective charge-carrier mode in the particle-hole channel, and the charge-carrier
normal self-energy (then the charge-carrier pseudogap) is obtained explicitly in
terms of
the spin excitation mode, while the electron normal self-energy (then the
normal-state
pseudogap) originated from the charge-carrier self-energy (then the charge-carrier
pseudogap) is due to the charge-spin recombination. Moreover, we have also shown
that (i) the spin excitation energy dispersion is anisotropically renormalized due
to the momentum dependence of the spin pseudogap. In particular, the antinodal spin
pseudogap effect is particularly notable in the slightly underdoped region, and
then this antinodal spin pseudogap rapidly decreases with the increase of doping in
the heavily underdoped region, eventually abrupt disappearing at around the optimal
doping; (ii) the electronic density of state at around the antinodal region is
gapped out by the normal-state pseudogap, and then the closed EFS contour is
truncated to a set of four disconnected Fermi arcs centered at around the nodal
region. By virtue of this full spin propagator and the reconstructed EFS, we have
investigated the low-temperature electrical transport in the underdoped pseudogap
phase of cuprate superconductors, where the scattering rate originated from the
umklapp scattering between electrons by the exchange of the full effective spin
propagator is derived within the framework of the Boltzmann transport theory. Our
results show that the dominant contribution to the low-temperature resistivity
mainly comes from the antinodal umklapp scattering. In particular, a {\it low
temperature} $T_{\rm scale}$ scales with $\Delta^{2}_{p}$ in the underdoped regime
due to the opening of the momentum dependence of the spin pseudogap. Moreover,
this $T_{\rm scale}$ as a function of doping presents a similar behavior of the
antinodal spin pseudogap crossover temperature $T^{*}_{\rm s}$, i.e.,
$T_{\rm scale}$ decreases with the increase of doping in the underdoped regime,
and then is reduced to a {\it very low temperature} in the overdoped regime, which
suggests a {\it possible correlation} between $T_{\rm scale}$ and the opening of
the antinodal spin pseudogap below $T^{*}_{\rm s}$. In the underdoped regime, the
resistivity exhibts a T-quadratic behaviour in the low-temperature region below
$T_{\rm scale}$, where the strength of the T-quadratic resistivity decreases with
the increase of doping. However, in the overdoped regime, the resistivity is
T-linear in the low-temperature region above $T_{\rm scale}$. The current results
together with the recent results\cite{Ma23} of the low-temperature T-linear
resistivity in the overdoped regime therefore show that (i) the electron umklapp
scattering from a spin excitation associated with the antinodes leads to the
T-linear resistivity in the weak coupling overdoped regime; (ii) as this electron
umklapp scattering flows to the strong coupling underdoped regime, the opening of
the momentum dependence of the spin pseudogap lowers the density of states of the
spin excitations at around the antinodal region in the response to the intense
umklapp scattering, which reduces the strength of the umklapp scattering from
electronic states into the antinodal region, and therefore leads to the
low-temperature T-quadratic form of the umklapp scattering rate. Concomitantly,
the low-temperature resistivity exhibits a dramatic switch from the T-linear
behaviour in the overdoped strange-metal phase to the T-quadratic behaviour in
the underdoped pseudogap phase.

It should be emphasized that the full effective spin propagator
$P({\bf k},{\bf p},{\bf k}',\omega)$ in Eq. (\ref{reduced-propagator}) is obtained
in the underdoped regime, where the AFLRO correlation is absent, although the AFSRO
correlation survives from the underdoped regime to the overdoped regime
\cite{Fujita12} as we have mentioned in subsection \ref{Boltzmann-theory}. In this
case, the umklapp scattering between electrons in Eq. (\ref{electron-collision-1})
by the exchange of this full effective spin propagator is better suited for the
description of the electrical transport in the underdoped regime. However, in the
{\it extremely light-doped regime} ($\delta \leq 0.05$), the AFLRO correlation has
been identified by INS, muon spin rotation, and other measurements
\cite{Kastner98,Fujita12,Vaknin87,Brewer88}. Moreover, the strength of the AFLRO
correlation as a function of the doping concentration has been established by NMR
measurements\cite{Kitaoka88}, where AFLRO is destructed by a few percent of the
doping concentration ($\sim 0.05$). In this extremely light-doped
regime, the electronic state of the system with some special features appear, in
particular, the low-temperature resistivity shows a upturn
\cite{Ando01,Ando04a,Ando04}. Although the mechanism causing this resistivity
upturn remains unclear, it is possible that this striking resistivity upturn
can be also described in terms of the umklapp scattering between electrons by the
exchange of the effective spin propagator, where the spin propagator should give a
suitable description of the special magnetic properties with the AFLRO correlation.
These and the related issues are under investigation now.

\section*{Acknowledgements}

This work is supported by the National Key Research and Development Program of
China under Grant Nos. 2023YFA1406500 and 2021YFA1401803, and the National Natural
Science Foundation of China (NSFC) under Grant Nos. 12274036 and 12247116. H.G.
acknowledge support from NSFC grant Nos. 11774019 and 12074022.

\begin{appendix}

\section{Derivation of full charge-carrier, full spin, and full electron
propagators} \label{charge-spin-propagator}

In this Appendix, the main goal is to derive the full charge-carrier propagator
$g({\bf k},\omega)$ in Eq. (\ref{FCCGF}), the full spin propagator
$D({\bf k},\omega)$ in Eq. (\ref{FSGF}), and the full electron propagator
$G({\bf k},\omega)$ in Eq. (\ref{EGF}) of the main text. Following the fermion-spin
transformation (\ref{CSSFS}), the $t$-$J$ model (\ref{tJ-model}) can be rewritten
as,
\begin{eqnarray}\label{cssham}
H&=&\sum_{\langle l\hat{\eta}\rangle}t(h^{\dagger}_{l+\hat{\eta}\uparrow}
h_{l\uparrow}S^{+}_{l}S^{-}_{l+\hat{\eta}}+h^{\dagger}_{l+\hat{\eta}\downarrow}
h_{l\downarrow}S^{-}_{l}S^{+}_{l+\hat{\eta}})\nonumber\\
&-&\sum_{\langle l\hat{\tau}\rangle}t'(h^{\dagger}_{l+\hat{\tau}\uparrow}
h_{l\uparrow}S^{+}_{l}S^{-}_{l+\hat{\tau}}+h^{\dagger}_{l+\hat{\tau}\downarrow}
h_{l\downarrow}S^{-}_{l}S^{+}_{l+\hat{\tau}})\nonumber\\
&-&\mu_{\rm h}\sum_{l\sigma}h^{\dagger}_{l\sigma}h_{l\sigma}+J_{\rm eff}
\sum_{\langle l\hat{\eta}\rangle}{\bf S}_{l}\cdot {\bf S}_{l+\hat{\eta}},
\end{eqnarray}
with $J_{{\rm eff}}=(1-\delta)^{2}J$, the charge-carrier doping concentration
$\delta=\langle h^{\dagger}_{l\sigma}h_{l\sigma}\rangle=\langle h^{\dagger}_{l}
h_{l}\rangle$, and the charge-carrier chemical potential $\mu_{\rm h}$. The above
$t$-$J$ model (\ref{cssham}) therefore describes a doped AF insulator as a sparse
density of the charge carriers moving in a background of an AF coupled square
lattice of spins, while the motion of the electrons rearranges the spin
configuration leading to the strong coupling between the charge and spin degrees
of freedom of the constrained electron.

\subsection{Full charge-carrier propagator}\label{full-charge-carrier-propagator}

In the early studies\cite{Feng15,Feng0306,Feng12,Feng15a}, it has been shown
that the interaction between the charge carriers directly from the kinetic energy
of the $t$-$J$ model (\ref{cssham}) by the exchange of the spin excitation
generates the charge-carrier pairing state in the particle-particle channel.
According to these early studies, the self-consistent equations that are
satisfied by the full charge-carrier diagonal and off-diagonal propagators in the
charge-carrier pairing state have been evaluated in terms of the Eliashberg
formalism\cite{Eliashberg60,Scalapino66}, and then in the charge-carrier
normal-state, these self-consistent equations are reduced as\cite{Feng15,Feng12},
\begin{eqnarray}
g({\bf k},\omega)=g^{(0)}({\bf k},\omega)+g^{(0)}({\bf k},\omega)
\Sigma^{({\rm h})}_{\rm ph}({\bf k},\omega)g({\bf k},\omega), ~~~~\label{HEDGF}
\end{eqnarray}
with the charge-carrier propagator of the $t$-$J$ model (\ref{cssham}) in the
MF approximation $g^{(0)-1}({\bf k},\omega)=\omega-\xi_{\bf k}$. From the above
self-consistent equation (\ref{HEDGF}), the full charge-carrier propagator can be
expressed explicitly as,
\begin{eqnarray}\label{FCCGF-A}
g({\bf k},\omega)={1\over\omega-\xi_{\bf k}
-\Sigma^{({\rm h})}_{\rm ph}({\bf k},\omega)},
\end{eqnarray}
which is the same as quoted in Eq. (\ref{FCCGF}) of the main text. Moreover, the
charge-carrier normal self-energy $\Sigma^{({\rm h})}_{\rm ph}({\bf k},\omega)$
has been derived as \cite{Feng15,Feng12},
%\begin{widetext}
\begin{eqnarray}\label{HESE}
\Sigma^{({\rm h})}_{\rm ph}({\bf k},i\omega_{n})&=&{1\over N}
\sum_{\bf p}{1\over\beta}\sum_{ip_{m}}g({\bf p}+{\bf k},ip_{m}+i\omega_{n})
\nonumber\\
&\times&P^{(0)}({\bf k},{\bf p},ip_{m}),~~~
\end{eqnarray}
%\end{widetext}
with the number of lattice sites N, the fermion and bosonic Matsubara frequencies
$\omega_{n}$ and $p_{m}$, respectively, and the MF effective spin propagator,
\begin{eqnarray}\label{ESP-1}
P^{(0)}({\bf k},{\bf p},\omega)={1\over N}\sum_{\bf q}
\Lambda^{2}_{{\bf p}+{\bf q}+{\bf k}}\Pi({\bf p},{\bf q},\omega),
\end{eqnarray}
where the MF spin bubble $\Pi({\bf p},{\bf q},\omega)$ is a convolution of two MF
spin propagators, and can be expressed as,
\begin{equation}\label{spin-bubble-1}
\Pi({\bf p},{\bf q},ip_{m})={1\over\beta}\sum_{iq_{m}}D^{(0)}({\bf q},iq_{m})
D^{(0)}({\bf q}+{\bf p},iq_{m}+ip_{m}),~~~
\end{equation}
with the bosonic Matsubara frequency $q_{m}$, and the MF spin propagator,
\begin{eqnarray}\label{SGF-1}
D^{(0)}({\bf k},\omega)={B_{\bf k}\over\omega^{2}-\omega^{2}_{\bf k}}
={B_{\bf k}\over 2\omega_{\bf k}}\left ( {1\over\omega-\omega_{\bf k}}
-{1\over\omega+\omega_{\bf k}}\right ).~~~~~
\end{eqnarray}
With the help of the above spin propagator (\ref{SGF-1}), the charge-carrier normal
self-energy $\Sigma^{({\rm h})}_{\rm ph}({\bf k},\omega)$ in Eq. (\ref{HESE}) can
be obtained as\cite{Feng15,Feng12},
\begin{eqnarray}\label{HESE-1}
\Sigma^{({\rm h})}_{\rm ph}({\bf k},\omega)&=&{1\over N^{2}}\sum_{{\bf pp}'\mu\nu}
(-1)^{\nu+1}Z^{({\rm h})}_{\rm F}\Omega_{{\bf p}{\bf p}'{\bf k}}\nonumber\\
&\times& {F^{({\rm h})}_{\mu\nu}({\bf p},{\bf p}',{\bf k})\over\omega
+(-1)^{\mu+1}\omega^{(\nu)}_{{\bf p}{\bf p}'}-\bar{\xi}_{{\bf p}+{\bf k}}},~~~
\end{eqnarray}
with $\mu~(\nu)=1,~2$, $\omega^{(\nu)}_{{\bf p}{\bf p}'}=\omega_{{\bf p}+{\bf p}'}
-(-1)^{\nu}\omega_{\bf p'}$, $\Omega_{{\bf p}{\bf p}'{\bf k}}=
\Lambda^{2}_{{\bf p}+{\bf p}'+{\bf k}}B_{{\bf p}'} B_{{\bf p}+{\bf p}'}
/[4\omega_{{\bf p}'}\omega_{{\bf p}+{\bf p}'}]$,
$F^{({\rm h})}_{\mu\nu}({\bf p},{\bf p}',{\bf k})
=n_{\rm F}[(-1)^{\mu+1}\bar{\xi}_{{\bf p}+{\bf k}}]
n^{(\nu)}_{{\rm 1B}{{\bf p}{\bf p}'}}+n^{(\nu)}_{{\rm 2B}{{\bf p}{\bf p}'}}$,
$n^{(\nu)}_{{\rm 1B}{{\bf p}{\bf p}'}}=1+n_{\rm B}(\omega_{{\bf p}'+{\bf p}})
+n_{\rm B}[(-1)^{\nu+1}\omega_{\bf p'}]$, and
$n^{(\nu)}_{{\rm 2B}{{\bf p}{\bf p}'}}
=n_{\rm B}(\omega_{{\bf p}'+{\bf p}})n_{\rm B}[(-1)^{\nu+1}\omega_{\bf p'}]$.

\subsection{Full spin propagator}\label{full-spin-propagator}

Starting from the $t$-$J$ model (\ref{cssham}) in the fermion-spin representation,
the full spin propagator in the normal-state has been evaluated as
\cite{Kuang15,Yuan01,Feng98},
\begin{eqnarray}\label{FSGF-A}
D({\bf k},\omega)&=&{1\over D^{(0)-1}({\bf k},\omega)
-\Sigma^{({\rm s})}_{\rm ph}({\bf k},\omega)},~~~~~
\end{eqnarray}
where the spin self-energy in the normal-state is derived in terms of the
collective charge-carrier mode in the particle-hole channel as,
\begin{eqnarray}\label{SSFN}
\Sigma^{({\rm s})}_{\rm ph}({\bf k},\omega)&=&-{1\over N^{2}}\sum_{\bf pq}
\Omega^{({\rm s})}_{{\bf k}{\bf p}{\bf q}}{F^{\rm (s)}({\bf k},{\bf p},{\bf q})
\over\omega^{2}-[\omega_{{\bf q}+{\bf k}}-(\bar{\xi}_{{\bf p}+{\bf q}}
-\bar{\xi}_{{\bf p}})]^{2}}, \nonumber\\
~~
\end{eqnarray}
with the vertex function $\Omega^{({\rm s})}_{{\bf k}{\bf p}{\bf q}}$ and the
function $F^{\rm (s)}({\bf k},{\bf p},{\bf q})$ that are given by,
\begin{subequations}\label{Functions-SSE}
\begin{eqnarray}
\Omega^{({\rm s})}_{{\bf k}{\bf p}{\bf q}}&=&{B_{{\bf q}+{\bf k}}\over
\omega_{{\bf q}+{\bf k}}}[Z^{({\rm h})}_{\rm F}]^{2}(\Lambda^{2}_{{\bf k}-{\bf p}}
+\Lambda^{2}_{{\bf p}+{\bf q}+{\bf k}}), ~~~~~~\\
F^{\rm (s)}({\bf k},{\bf p},{\bf q})&=&[\omega_{{\bf q}+{\bf k}}
-(\bar{\xi}_{{\bf p}+{\bf q}}-\bar{\xi}_{{\bf p}})]
\{n_{\rm B}(\omega_{{\bf q}+{\bf k}})[n_{\rm F}(\bar{\xi}_{{\bf p}})\nonumber\\
&-&n_{\rm F}(\bar{\xi}_{{\bf p}+{\bf q}})]-[1-n_{\rm F}(\bar{\xi}_{{\bf p}})]
n_{\rm F}(\bar{\xi}_{{\bf p}+{\bf q}})\},~~~~~~~~
\end{eqnarray}
\end{subequations}
respectively. Substituting the above spin self-energy in Eq. (\ref{SSFN}) into
Eq. (\ref{FSGF-A}), the full spin propagator in the normal-state can be expressed
as,
\begin{eqnarray}\label{FSGF-AA}
D({\bf k},\omega)&=&{B_{\bf k}\over \omega^{2}-\omega^{2}_{\bf k}
-B_{\bf k}\Sigma^{({\rm s})}_{\rm ph}({\bf k},\omega)},~~~~~
\end{eqnarray}
which is the same as quoted in Eq. (\ref{FSGF}) of the main text.

\subsection{Full electron propagator}\label{full-electron-propagator}

In order to obtain the electron propagator, the full charge-spin recombination
scheme has been developed\cite{Feng15a}, where it has been shown that the coupling
form between the electron and spin excitation in the $t$-$J$ model in the
normal-state is the same as that between the charge carrier and spin excitation in
Eq. (\ref{HEDGF}), and then the self-consistent equation satisfied by the full
electron propagator in the normal-state can be derived directly as\cite{Feng15a},
\begin{eqnarray}
G({\bf k},\omega)=G^{(0)}({\bf k},\omega)+G^{(0)}({\bf k},\omega)
\Sigma_{\rm ph}({\bf k},\omega)G({\bf k},\omega), ~~~~\label{EDGF}
\end{eqnarray}
where $G^{(0)-1}({\bf k},\omega)=\omega-\varepsilon_{\bf k}$ is the electron
propagator of the $t$-$J$ model (\ref{tJ-model}) in the tight-binding
approximation. From the above equation (\ref{EDGF}), the full electron propagator
$G({\bf k},\omega)$ can be expressed explicitly as,
\begin{eqnarray}\label{EGF-A}
G({\bf k},\omega)={1\over \omega-\varepsilon_{\bf k}
-\Sigma_{\rm ph}({\bf k},\omega)},
\end{eqnarray}
which is the same as quoted in Eq. (\ref{EGF}) of the main text. In particular,
the electron normal self-energy,
%\begin{widetext}
\begin{eqnarray}\label{ESE}
\Sigma_{\rm ph}({\bf k},i\omega_{n})&=&{1\over N}\sum_{\bf p}{1\over\beta}
\sum_{ip_{m}}G({\bf p}+{\bf k},ip_{m}+i\omega_{n})\nonumber\\
&\times&P^{(0)}({\bf k},{\bf p},ip_{m}),~~~
\end{eqnarray}
%\end{widetext}
has been evaluated as\cite{Feng15a},
\begin{eqnarray}\label{ESE-5}
\Sigma_{\rm ph}({\bf k},\omega)&=&{1\over N^{2}}\sum_{{\bf pp'}\mu\nu}(-1)^{\nu+1}
Z_{\rm F}\Omega_{\bf pp'k}\nonumber\\
&\times& {F_{\mu\nu}({\bf p},{\bf p}'{\bf k})\over\omega+(-1)^{\mu+1}
\omega_{\nu{\bf p}{\bf p}'}- \bar{\varepsilon}_{{\bf p}+{\bf k}}},~~~
\end{eqnarray}
with the function,
\begin{eqnarray}
F_{\mu\nu}({\bf p},{\bf p}'{\bf k})=n_{\rm F}[(-1)^{\mu+1}
\bar{\varepsilon}_{{\bf p}+{\bf k}}]n^{(\nu)}_{{\rm 1B}{\bf pp'}}
+n^{(\nu)}_{{\rm 2B}{\bf pp'}},~~~~~
\end{eqnarray}
where $n^{(\nu)}_{{\rm 1B}{\bf pp'}}=1+n_{\rm B}(\omega_{{\bf p}'+{\bf p}})
+n_{\rm B}[(-1)^{\nu+1}\omega_{\bf p'}]$, and $n^{(\nu)}_{{\rm 2B}{\bf pp'}}
=n_{\rm B}(\omega_{{\bf p}'+{\bf p}})n_{\rm B}[(-1)^{\nu+1}\omega_{\bf p'}]$. This
electron normal self-energy characterizes a competition between the kinetic energy
and magnetic energy in the $t$-$J$ model (\ref{tJ-model}).

\end{appendix}

\end{document}